\documentclass[aps,pra,notitlepage,superscriptaddress,floatfix,10pt,nofootinbib,twocolumn]{revtex4-2}
\usepackage[utf8]{inputenc}
\usepackage[english]{babel}
\usepackage[T1]{fontenc}
\usepackage{amsmath}

\usepackage{tikz}
\usepackage{lipsum}
\usepackage{graphicx}% Include figure files
\usepackage{bbm}
\usepackage[
            colorlinks=true,
            linkcolor=blue,
            citecolor=blue,
            urlcolor=blue
            ]{hyperref}

\newcommand{\ketbra}[2]{|#1\rangle\langle #2|}
\newcommand{\ket}[1]{|#1\rangle}
\newcommand{\bra}[1]{\langle #1|}
\newcommand{\va}{\vec{a}}

\newcommand{\vx}{\vec{x}}
\newcommand{\vi}{\vec{i}}
\newcommand{\vt}{\vec{t}}
\newcommand{\ve}{\vec{e}}
\newcommand{\vz}{\vec{z}}

\newcommand{\oi}{\overline{i}}
\newcommand{\oj}{\overline{j}}
\newcommand{\ta}{\theta}
\newcommand{\vta}{\vartheta}

\newcommand{\sA}{\textsf{A}}
\newcommand{\sB}{\textsf{B}}

\newcommand{\cB}{\mathcal{B}}
\newcommand{\cN}{\mathcal{N}}
\newcommand{\cS}{\mathcal{S}}

\newtheorem{definition}{Definition}

\begin{document}

\title{Hierarchical certification of nonclassical network correlations}

\author{Ming-Xing Luo}
\email{mxluo@swjtu.edu.cn}
\affiliation{Information Security and National Computing Grid Laboratory, Southwest Jiaotong University, Chengdu 610031, China}
\affiliation{CAS Center for Excellence in Quantum Information and Quantum Physics, Hefei, 230026, China}
\author{Xue Yang}
\affiliation{Information Security and National Computing Grid Laboratory, Southwest Jiaotong University, Chengdu 610031, China}

\author{Alejandro Pozas-Kerstjens}
\email{physics@alexpozas.com}
\affiliation{Group of Applied Physics, University of Geneva, 1211 Geneva 4, Switzerland}
\affiliation{Constructor University, Geneva, Switzerland}
\affiliation{Instituto de Ciencias Matem\'aticas (CSIC-UAM-UC3M-UCM), 28049 Madrid, Spain}

\begin{abstract}
With the increased availability of quantum technological devices, it becomes more important to have tools to guarantee their correct nonclassical behavior.
This is especially important for quantum networks, which constitute the platforms where multipartite cryptographic protocols will be implemented, and where guarantees of nonclassicality translate into security proofs.
We derive linear and nonlinear Bell-like inequalities for networks, whose violation certifies the absence of a minimum number of classical sources in them.
We do so, first, without assuming that nature is ultimately governed by quantum mechanics, providing a hierarchy interpolating between network nonlocality and full network nonlocality.
Second we insert this assumption, which leads to results more amenable to certification in experiments.
\end{abstract}

\maketitle

%%%%%%%%%%%%%%%%%%%%%%%%%%%%%%%%%%%%%%%%%%%%%%%%%%%%%%%%%%%%%%%%%%%%%%%%%%%%%

\section{Introduction}
Bell's theorem asserts that models based on classical shared randomness cannot replicate the probabilistic predictions of quantum theory in all circumstances \cite{Bell}.
In particular, certain correlations generated from local measurements on entangled quantum systems \cite{EPR} cannot be simulated by local hidden variable models.
Bell nonlocality, which can be generated from any bipartite entangled system regardless of its dimension \cite{Gisin1991}, represents a type of quantumness that is significant both in the fundamental theory and in various applications \cite{Mayers,BCPS,Ekert,Acin2007,Pironio2010}.
The methods for detecting nonlocality can be readily extended to multipartite scenarios by separating the parties along bipartitions \cite{Svetlichny,Gallego,Bancal}.
Traditionally, the parties within a bipartition are allowed to communicate with each other, and genuine multipartite nonlocality is then defined as the impossibility to write a multipartite probability distribution with a local model along any possible bipartition of the parties.

\begin{figure}
    \centering
    \includegraphics[width=0.95\columnwidth]{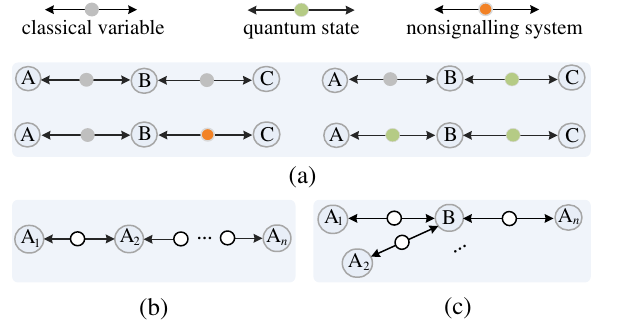}
    \caption{The main networks discussed in this work.
    (a) The network underlying simple entanglement swapping experiments. Correlations generated in the network are considered NN if they cannot be generated as in the top-left setup, FNN if they cannot be generated as in the bottom-left setup, and FQNN if they cannot be generated as in the top-right setup. In the appendices we show that realizations in the bottom-right setup produce all these types of correlations.
    (b) The generalization of the previous scenario to chains, and (c) its generalization to stars.
    }
    \label{fig-1}
\end{figure}

A recent trend consists of investigating the correlations that can be generated in network structures where, instead of considering a global joint system being shared between the parties, several independent sources distribute systems to different collections of them.
For example, quantum networks use sources of entangled systems, as depicted in Fig.~\ref{fig-1}, providing a scalable framework for constructing the quantum internet and implementing diverse quantum applications \cite{ACL,Kimble,SSdG,Wehner,Renou2021}.
Quantum multipartite nonlocality has been investigated in these scenarios \cite{Popescu,BGP,Fritz,Chave2015,Allen2017,Aberg,networkReview}, leading to new fundamental insights on quantum theory \cite{Renou2021,Abiuso2021}.
Nevertheless, exploring nonlocal correlations in networks implies considerable challenges since the relevant sets of correlations are not convex, and thus objects such as linear Bell inequalities cannot characterize them completely.

Nonlinear Bell inequalities have been used to study network nonlocality (NN), especially in the tripartite network underlying entanglement swapping depicted in Fig.~\ref{fig-1}(a) \cite{BGP,BRGP,TSCA,BBBC}.
These inequalities have been extended to more general networks with more parties and sources of physical systems (\cite{Chave2016,RBBP,GMTR,Luo2018,Tavaki2021}, see also the review \cite{networkReview}).
However, in analogy with the need to define genuine multipartite nonlocality as a means to guarantee global nonclassical phenomena, the observation of network nonlocality only guarantees nonclassical behavior in some part of the network.
It is possible to guarantee a complete absence of classical components via observing genuine multipartite nonlocality, but it is known that accounting for the network structure allows for milder requirements in experimentally relevant parameters, such as detection efficiency \cite{Pozas2019}.
Currently, there exist two complementary definitions of genuinely nonlocal behaviors in networks: genuine network nonlocality \cite{Supic2022} and full network nonlocality (FNN) \cite{Alej2022}.
The latter uses minimal assumptions, and has been recently observed in experimental implementations of different network structures \cite{Haakansson2022,Huang2022,Wang2023,Gu2023}.

In this work, we study fully network nonlocal correlations under the assumption that physical systems are ultimately governed by quantum mechanics.
We derive new Bell-like inequalities, both linear and nonlinear, for star and chain networks (see Figs.~\ref{fig-1}(b) and (c)).
The violation of these inequalities certifies that the corresponding statistics could not have been produced had there been at least a number of classical sources in the network.
In many cases, we are able to provide a single inequality that guarantees that no source in the network distributes classical physical systems, in contrast with previous works \cite{Alej2022,Wang2023} that needed one inequality per source in the network.
In other cases, we provide bounds for each number of classical sources in the network.
In addition to this, we provide FNN bounds for the inequalities, and demonstrate violations of both types of bounds within quantum mechanics.
Finally, we discuss how these results can be exploited to guarantee nonclassicality of the sources of arbitrary networks.

\section{Models for non-full correlations}
Nonlocality in networks is defined in an analogous manner to the bipartite case \cite{Bell,BCPS}, namely by opposition to having a local model.
This is, network nonlocal correlations do not admit a decomposition in the following form
\begin{equation}
    P(\va|\vx)
    =\int_{\Omega}\prod_{j=1}^md\mu(\lambda_j)\prod_{i=1}^{n} P(a_i|x_i,\bar{\lambda}_{i}),
    \label{eqn-2}
\end{equation}
where $\va=a_1\dots{}a_n$ denotes the parties' outputs, $\vec{x}=x_1\dots{}x_n$ their inputs, $\lambda_j$ the classical variable distributed by source $j$, and $\bar{\lambda}_i$ the set of variables that arrive to party $i$, i.e., the set of classical physical systems that arrive to party $i$ from the sources that connect to it.
As an illustration, take the network known as the bilocality scenario, depicted in Fig.~\ref{fig-1}(a).
Network-local correlations in this network (i.e., bilocal correlations) admit a description of the form \cite{BGP,BRGP}
\begin{align*}
    P(a,b,c|x,y,z) = \int&\text{d}\mu(\lambda_1)\text{d}\mu(\lambda_2) \\
    &P(a|x,\lambda_1) P(b|y,\lambda_1,\lambda_2) P(c|z,\lambda_2).
\end{align*}
When $m=1$, Eq.~\eqref{eqn-2} recovers the standard notion of a local hidden variable model \cite{BCPS}.

Many works have focused on detecting when multipartite correlations do and do not admit realizations in the form of Eq.~\eqref{eqn-2}, producing many Bell-like inequalities \cite{BGP,BRGP,Chave2016,RBBP,GMTR,Luo2018,Tavaki2021,networkReview}.
However, in generalizing the standard notion of a Bell inequality, inequalities tailored for detecting network nonlocality can be violated even when just one of the sources in the network is nonclassical \cite{Fritz}.
This motivated the development of new notions beyond network nonlocality whose observation guarantees stronger quantum properties, for instance the absence of sources of classical systems \cite{Alej2022} or the presence of joint quantum measurements \cite{Supic2022} in the network.
Concretely, Ref.~\cite{Alej2022} defines as ``not interesting'' correlations (i.e., the analogous of Eq.~\eqref{eqn-2} in the setting of network nonlocality) those that can be generated if and only if there is at least one source in the network that distributes classical physical systems.
Correlations that cannot be generated in this way are called FNN.

In this work we consider two refinements of the definition of FNN in Ref.~\cite{Alej2022}.
First, we define a hierarchy of levels between NN and FNN, determined by the number of classical sources that can be used for reproducing the correlation.

\begin{definition}[$\ell$-level network nonlocality]
    A multipartite probability distribution is $\ell$-level network nonlocal ($\ell$-NN) relative to a network if it cannot be generated when at least $\ell$ sources in the network distribute classical physical systems, and the rest are allowed to distribute physical systems only limited by no-signaling. Otherwise, the distribution is $\ell$-level network local ($\ell$-NL).
    \label{def:lNL}
\end{definition}

It is clear that the above definition coincides with that of NN when $\ell$ equals the number of sources in the network (i.e., when $\ell=m$ in Eq.~\eqref{eqn-2}), and with FNN when $\ell=1$.
As an illustration, consider the star network in Fig.~\ref{fig-1}(c) with three branch parties, $\{\sA_1,\sA_2,\sA_3\}$.
$\ell$-NL correlations in this network admit distributions of the forms
\begin{align}
    P^\text{2-NL}(&a_1,a_2,a_3,b|x_1,x_2,x_3,y)= \int\text{d}\mu(\lambda_1)\text{d}\mu(\lambda_2) &\notag\\
    & P(a_1|x_1,\lambda_1)P(a_2|x_2,\lambda_2)P_\text{NSI}(a_3,b|x_3,y,\lambda_1,\lambda_2),& \label{eq:2NL}\\
    P^\text{1-NL}(&a_1,a_2,a_3,b|x_1,x_2,x_3,y)= \notag\\
    \int\text{d}&\mu(\lambda_1) P(a_1|x_1,\lambda_1)P_\text{NSI}(a_2,a_3,b|x_2,x_3,y,\lambda_1),\label{eq:1NL}
\end{align}
or of analogs under permutations of the branch parties.
$P_\text{NSI}$ are distributions only constrained by the principles of no-signaling and the independence of the sources \cite{Gisin2020}.
This means that for the distribution in Eq.~\eqref{eq:2NL} we have that $\sum_{a_3}P_\text{NSI}(a_3,b|x_3,y,\lambda_1,\lambda_2)=P_\text{NSI}(b|y,\lambda_1,\lambda_2)$ and $\sum_{b}P_\text{NSI}(a_3,b|x_3,y,\lambda_1,\lambda_2)=P(a_3|x_3)$, and  for the case of the distribution in Eq.~\eqref{eq:1NL} we require that $\sum_{a_i}P_\text{NSI}(a_2,a_3,b|x_2,x_3,y,\lambda_1)=P_\text{NSI}(a_{j\not=i},b|x_{j\not=i},y,\lambda_1)$ and $\sum_{b}P_\text{NSI}(a_2,a_3,b|x_2,x_3,y,\lambda_1)=P(a_2|x_2)P(a_3|x_3)$.

Definition \ref{def:lNL} allows to make theory-independent statements on the presence of classical sources in networks \cite{Wang2023}.
However, inserting the additional assumption that the nonclassical sources in the network distribute systems that obey quantum theory is a natural one that, as we will show, leads to milder requirements on the experimentally relevant parameters (such as the visibility under white noise) needed to observe nonlocality.
Thus, we also study the hierarchy in Definition~\ref{def:lNL} under this assumption.
This leads to an analogous hierarchy, and to the definition of fully quantum network nonlocality (FQNN):

\begin{definition}[$\ell$-level quantum network nonlocality]
    A multipartite probability distribution is $\ell$-level quantum network nonlocal ($\ell$-QNN) relative to a network if it cannot be generated when at least $\ell$ sources in the network distribute classical physical systems, and the rest are allowed to distribute arbitrary multipartite quantum systems. Otherwise, the distribution is $\ell$-level quantum network local ($\ell$-QNL). We say that a correlation is fully quantum network nonlocal if it is 1-QNN.
    \label{def:lQNL}
\end{definition}

\begin{figure}
    \centering
    \includegraphics[width=0.95\columnwidth]{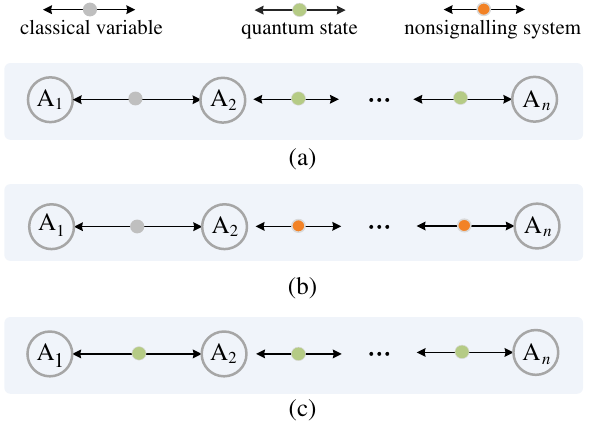}
    \caption{The $n$-partite chain network. Correlations that do not admit a realization neither in the form of panel (a) nor with any other positioning of the classical variable are FQNN, and those that do not admit a realization neither in the form of panel (b) nor with any other positioning of the classical variable are FNN. There exist correlations of both types that can be produced in the setup depicted in panel (c).}
    \label{fig:chains}
\end{figure}

Going back to the example of the star network with three branch parties, the corresponding distributions $P^\text{2-QNL}$ and $P^\text{1-QNL}$ have the same form as Eqs.~\eqref{eq:2NL} and \eqref{eq:1NL}, where now the distributions $P_\text{NSI}$ are instead distributions that follow the corresponding Born's rule, i.e. $\text{Tr}(\Pi_{a_3}^{x_3}\otimes\Pi_b^{y,\lambda_1,\lambda_2}\cdot\rho_{A_3B})$ for Eq.~\eqref{eq:2NL} and $\text{Tr}(\Pi_{a_2}^{x_2}\otimes\Pi_{a_3}^{x_3}\otimes\Pi_b^{y,\lambda_1}\cdot\rho_{A_2B}\otimes\rho_{A_3B})$ for Eq.~\eqref{eq:1NL}.

In the remainder of this manuscript we analytically prove inequalities, both linear and nonlinear, satisfied by $\ell$-NL and $\ell$-QNL correlations in large families of networks and for several values of $\ell$, including complete hierarchies (i.e., $\ell\in[1,\dots,m]$) for some scenarios. We must note that, in general, obtaining analytical results for $\ell$-NL is hard. This is due to the lack of a formal description of all generalized states and effects that lead to nonsignaling distributions. For a more systematic analysis than the one done in this work one could use inflation methods (see, e.g., \cite{Alej2022,Wang2023,Boghiu2023} for examples of its use in the derivation of witnesses of FNN), to create tractable relaxations of the sets of $\ell$-NL distributions.

\section{Nonlinear inequalities for chain networks}\label{chainna}
The first family of networks that we study is that of chain networks, depicted in Figs.~\ref{fig-1}(b) and \ref{fig:chains}.
Chain networks are comprised of sources of bipartite states with the parties arranged in the form of a chain, such that with the exception of the two, extremal parties, all the remaining ones receive systems from two different sources. We will initially analyze the simplest case of the tripartite chain network, and extend our results to longer chains in a second step.

\subsection{The tripartite chain}
Let us begin with the tripartite chain network depicted in Fig.~\ref{fig-1}(a), also known as the bilocality network \cite{BGP,BRGP}. It comprises two bipartite sources, that distribute systems to three parties, in such a way that one of the parties receives a system from each of the sources.
This is the simplest non-trivial network and thus has received substantial attention in the literature \cite{BGP,BRGP,TSCA,BBBC,Luo2018}.
When the parties have binary inputs and outputs, the inequality derived in Ref.~\cite{BGP}, commonly known as the bilocality inequality or the $I$-$J$ inequality, holds for bilocal correlations.
This is, that correlations that admit an expression in terms of Eq.~\eqref{eqn-2} for the bilocality network satisfy
\begin{equation}
    \sqrt{|I|}+\sqrt{|J|} \leq 1,
    \label{eq:BGP}
\end{equation}
where $I$ and $J$ are defined by $I=\frac{1}{4}\left\langle A_0B^0C_0+A_1B^0C_0+A_0B^0C_1+A_1B^0C_1\right\rangle$ and $J=\frac{1}{4}\left\langle A_0B^1C_0-A_1B^1C_0-A_0B^1C_1+A_1B^1C_1\right\rangle$, $A_i$ and $C_k$ are the (dichotomic) measurement operators of the extreme parties, $B$ denotes one measurement with two-bit outcomes, depicted $b^0b^1$, and the correlator $\left\langle A_{x}B^yC_{z}\right\rangle$ is given by $\sum_{a,b^0b^1,c}(-1)^{a+b^y+c}P(a,b^0b^1, c|x,z)$.

In this section we will compute bounds on the left-hand side of Eq.~\eqref{eq:BGP} under the models considered in Definitions \ref{def:lNL} and \ref{def:lQNL}.
The case of $\ell=2$ is, as discussed before, the standard notion of network nonlocality given by Eq.~\eqref{eqn-2}.
Let us then consider the case of $\ell=1$, i.e., when only one of the sources distributes classical systems.
Without loss of generality, let us assume that the source distributing classical systems is that between Alice and Bob.
This implies that
\begin{widetext}
    \begin{align}
        \sqrt{|I|}+\sqrt{|J|}=\,&\frac{1}{2}\sqrt{|\left\langle A_0B^0C_0+A_1B^0C_0+A_0B^0C_1+A_1B^0C_1\right\rangle|} +\frac{1}{2}\sqrt{|\left\langle A_0B^1C_0-A_1B^1C_0-A_0B^1C_1+A_1B^1C_1\right\rangle|} \nonumber \\
        \leq\, &\frac{1}{2}\sqrt{|\left\langle A_0+A_1\right\rangle| |\left\langle B^0C_0+B^0C_1\right\rangle|} +\frac{1}{2}\sqrt{|\left\langle A_0-A_1\right\rangle| |\left\langle B^1C_0-B^1C_1\right\rangle|}.
        \label{A001}
    \end{align}
\end{widetext}
Now, assume that $\left\langle A_0+A_1\right\rangle\geq0$ and define $x=\left\langle A_0+A_1\right\rangle-1$, so that $x\in [-1,1]$.
Note that $|\left\langle A_0-A_1\right\rangle|+|\left\langle A_0+A_1\right\rangle|\leq 2\max\{|\left\langle A_0\right\rangle|,|\left\langle A_1\right\rangle|\}\leq 2$.
Therefore, we have that $|\left\langle A_0-A_1\right\rangle|\leq 2 - |\left\langle A_0 + A_1\right\rangle|= 1-x$.
Using this, Eq.~\eqref{A001} reads
\begin{align}
   \sqrt{|I|}+\sqrt{|J|} \leq\, &\frac{1}{2}\left(\sqrt{(1+x) |\left\langle B^0C_0+B^0C_1\right\rangle|}\right.\notag\\
   &\left.\quad+\sqrt{(1-x) |\left\langle B^1C_0-B^1C_1\right\rangle|}\right) \nonumber \\
    =\,&\frac{1}{\sqrt{2}}\left(\cos\theta \sqrt{|\left\langle B^0C_0+B^0C_1\right\rangle|}\right.\notag\\
    &\left.\qquad + \sin\theta\sqrt{|\left\langle B^1C_0-B^1C_1\right\rangle|}\right)
    \label{A002}
    \\
    &\hspace*{-1cm}\leq\, \frac{1}{\sqrt{2}}\sqrt{|\left\langle B^0C_0+B^0C_1\right\rangle+|\left\langle B^1C_0-B^1C_1\right\rangle|},
    \label{A003}
\end{align}
where in Eq.~(\ref{A002}) we have rewritten $1+x:=2\cos^2\theta$ and $1-x:=2\sin^2\theta$, and in Eq.~(\ref{A003}) we have used that $a\cos\theta + b\sin\theta\leq\sqrt{a^2+b^2}$.
If, instead, $\left\langle A_0+A_1\right\rangle<0$, one can define $x=\left\langle A_0+A_1\right\rangle+1$ and the same result follows.

Next, note that the object $B^y$ only considers the $y$-th bit of Bob's output.
One can thus view the correlators $\left\langle B^yC_z\right\rangle$ as generated by the following simulated, two-input, two-output distribution:
\begin{equation*}
   P_{\text{sim}}(b,c|y,z)= P(b=b^y,c|z) = \sum_{b^0,b^1} \delta_{b,b^y}P(b^0b^1,c|z),
\end{equation*}
which is obtained by Bob outputting only the $y$-th bit upon choosing measurement $y$.
Moreover, it is easy to see that $|\left\langle B^0C_0+B^0C_1\right\rangle|+|\left\langle B^1C_0-B^1C_1\right\rangle|$ is the largest of $|\left\langle B^0C_0+B^0C_1+(B^1C_0-B^1C_1)\right\rangle|$ and $|\left\langle B^0C_0+B^0C_1-(B^1C_0-B^1C_1)\right\rangle|$.
These are two variants of the CHSH operator under relabeling of Charlie's input.
Thus, it follows that
\begin{eqnarray*}
    \sqrt{|I|}+\sqrt{|J|}\leq \frac{1}{\sqrt{2}}\sqrt{\text{CHSH}(B,C)}.
\end{eqnarray*}

Therefore, the bounds on $\sqrt{|I|}+\sqrt{|J|}$ for FNN and FQNN are connected to the maximum value of the CHSH inequality \cite{CHSH} that Bob and Charlie can achieve.
When Bob and Charlie share a classical physical system, $\text{CHSH}(B,C)\leq 2$ and thus any value $\sqrt{|I|}+\sqrt{|J|}> \sqrt{2}/\sqrt{2}=1$ implies standard network nonlocality, as derived in Refs.~\cite{BGP,BRGP}.
If they share an arbitrary system only limited by the no-signaling principle, then $\text{CHSH}(B,C)\leq 4$, and any value $\sqrt{|I|}+\sqrt{|J|}> \sqrt{4}/\sqrt{2}=\sqrt{2}$ implies FNN, as derived in Ref.~\cite{Alej2022}.
Finally, if the parties share a quantum system, then $\text{CHSH}(B,C)\leq 2\sqrt{2}$ and any value $\sqrt{|I|}+\sqrt{|J|}> \sqrt{2\sqrt{2}}/\sqrt{2}=2^{1/4}$ implies FQNN.
This is a proof that the bound first provided in Ref.~\cite{SBBP} is tight.

The same argumentation can be done for the case where the classical source is shared between Bob and Charlie, arriving to the same inequality, $\sqrt{|I|}+\sqrt{|J|}\leq2^{1/4}$. Therefore, this is a single inequality whose violation guarantees that none of the sources in the scenario is classical.
This is in contrast to earlier works on detection of FNN \cite{Alej2022,Wang2023}, which needed to violate as many inequalities as sources in the network in order to guarantee that all of them were nonclassical.
Moreover, both bounds are tight: the bound for FQNN is saturated by the strategy in Ref.~\cite{SBBP} and the bound for FNN is saturated by the strategy in Ref.~\cite{Alej2022}.
In fact, the bound for FNN coincides with the algebraic maximum of $\sqrt{|I|}+\sqrt{|J|}$, computed in Ref.~\cite{Alej2022}.
This means that $\sqrt{|I|}+\sqrt{|J|}$ cannot be used for detecting FNN, but it can be used for detecting FQNN.
Indeed, take the sources to distribute pure states, $\rho_i=\ketbra{\phi_i}{\phi_i}$, given by
\begin{equation}
    \ket{\phi_i}=\cos\ta_i\ket{00}+\sin\ta_i\ket{11},
    \label{A006}
\end{equation}
where $\ta_i\in (0,\frac{\pi}{2})$.
Let also $A_x=\cos\vta \sigma_3+(-1)^x\sin\vta \sigma_1$ and likewise for $C_z$, and $B^{b^0b^1}\in \{\Phi_{0}, \Phi_1, \Psi_0, \Psi_1\}$, where $\sigma_1$ and $\sigma_3$ are the Pauli matrices, and $\rho=\ket{\rho}\bra{\rho}$ with $\ket{\Phi_i}=(\ket{00}+(-1)^i\ket{11})/\sqrt{2}$ and $\ket{\Psi_i}=(\ket{01}+(-1)^i\ket{10})/\sqrt{2}$ being the Bell states.
A straightforward evaluation leads to
\begin{eqnarray*}
    I=\cos^2\vta,\qquad J=\sin^2\vta \sin2\ta_1\sin2\ta_2,
\end{eqnarray*}
and we thus have that 
\begin{equation*}
    \sqrt{|I|}+\sqrt{|J|}=\cos\vta+\sin\vta \sqrt{\sin2\ta_1\sin2\ta_2}.
\end{equation*}
This quantity exceeds $2^{1/4}$, for instance, for any $\ta_i$s satisfying $\sin2\ta_1\sin2\ta_2\geq \sqrt{2}-1\approx 0.4142$, whenever choosing $\vta$ such that $\cos\vta=1/\sqrt{1+\sin2\ta_1\sin2\ta_2}$.
Moreover, the maximum value over all possible $\ta_1$, $\ta_2$, $\vta$ is $\sqrt{2}$, which is the algebraic maximum.

\subsection{Long chains}\label{sec:chainnl:long}
The bilocality inequality can be extended to chain networks of arbitrary length by using the notion of $k$-independent sets of parties \cite{Luo2018},\footnote{As defined in Ref.~\cite{Luo2018}, $k$-independence indicates that the maximum number of parties in the network that share no source with each other is $k$.} and with it building appropriate operators $I$ and $J$ in order to construct Bell-like inequalities whose violations witness $\ell$-NN and $\ell$-QNN. Indeed, Ref.~\cite{Luo2018} showed that a suitable generalization of the $I$ and $J$ operators is
\begin{equation*}
    \begin{aligned}
        I&=\left\langle\Pi_{{\rm{}odd}\, i }A^+_{x_{i}}\Pi_{{\rm{}even}\, j}A_{x_{j}=0}\right\rangle,\\
        J&=\left\langle\Pi_{{\rm{}odd}\, i} A^{-}_{x_{i}}\Pi_{{\rm{}even}\,j}A_{x_{j}=1}\right\rangle,
    \end{aligned}
\end{equation*}
with $A^\pm_{x_{i}}=(A_{x_{i}=0}\pm A_{x_{i}=1})/2$, and the correlators representing the combination of probabilities given by $\left\langle A_{x_{1}}\cdots A_{x_{n}}\right\rangle=\sum_{\vec{a}}(-1)^{\sum_{i=1}^na_{i}}P(\vec{a}|\vec{x})$.
Reference~\cite{Luo2018} proved that for any $k$-independent network the correlations generated by classical sources satisfy the Bell inequality $|I|^{\frac{1}{k}}+|J|^{\frac{1}{k}}\leq 1$, and that this inequality admits quantum violations using maximally entangled and GHZ states.

In the following we analyze the case where the network contains an odd number of parties and the nonclassical sources are required to follow quantum theory.
In other words, we restrict to the case of $\ell$-QNN in chain networks with an odd number of parties.
We defer the analysis of the remaining cases, namely calculations for $\ell$-NN and for chain networks with an even number of parties, to Appendix~\ref{app:nlchain}.
We furthermore distinguish two sub-cases, namely when the network contains $\ell\geq (n-1)/2$ classical sources, and when it contains $\ell < (n-1)/2$ classical sources.

\textbf{Case} $\mathbf{\ell\geq (n-1)/2}$:
In this case, there are more classical than quantum sources.
This implies that there are no more than $(n-1)/2$ non-adjacent pairs of parties $(\sA_i,\sA_{i+1})$ who share a quantum system, and thus, the network is a $k$-independent network \cite{Luo2018} with $k\leq (n-1)/2$.
Let $S$ consist of all odd integers, $i$, such that $\sA_i$ receives only classical sources.
The fact that there are more classical than quantum sources implies that $|S|\geq 1$, and from the classicality of the measurements of $\sA_i$ when $i\in S$ it follows that
\begin{align}
    I^{\frac{2}{n+1}}&+J^{\frac{2}{n+1}} =\left\langle\prod\limits_{{\rm{}odd}\, i }A^+_{x_{i}}\prod\limits_{{\rm{}even}\, j}A_{x_{j}=0}\right\rangle^{\frac{2}{n+1}}
    \notag\\
    &\qquad\qquad\quad+\left\langle\prod\limits_{{\rm{}odd}\, i }A^{-}_{x_{i}}\prod\limits_{{\rm{}even}\, j}A_{x_{j}=1}\right\rangle^{\frac{2}{n+1}} \notag\\
    &\leq \left\langle\prod\limits_{i\in S}A^{+}_{x_{i}}\right\rangle^{\frac{2}{n+1}}\left\langle\prod\limits_{{\rm{}odd}\, i; i\not\in S}A^+_{x_{i}}\prod\limits_{{\rm even}\, j}A_{x_{j}=0}\right\rangle^{\frac{2}{n+1}} \notag\\
    \quad\quad&\quad+\left\langle\prod\limits_{i\in S}A^{-}_{x_{i}}\right\rangle^{\frac{2}{n+1}}\left\langle\prod\limits_{{\rm{}odd}\, i; i\not\in S}A^{-}_{x_{i}}\prod\limits_{{\rm{}even}\, j}A_{x_{j}=1}\right\rangle^{\frac{2}{n+1}}\!\!\!\!\!,
    \label{DD1}
\end{align}
where, recall, $A^\pm_{x_i}=\frac12(A_{x_i=0}\pm A_{x_i=1})$.

Reference \cite{Luo2018} gave a quantum bound in the $k$-independent network, which when $k=\frac{n-1}{2}$ reads:
\begin{align*}
    &\left\langle\prod\limits_{{\rm odd}\, i; i\not\in S}A^+_{x_{i}}\prod\limits_{{\rm even}\, j}A_{x_{j}=0}\right\rangle^{\frac{2}{n-1}}
    \\
    &+\left\langle\prod\limits_{{\rm odd}\, i; i\not\in S}A^-_{x_{i}}\prod\limits_{{\rm even}\, j}A_{x_{j}=1}\right\rangle^{\frac{2}{n-1}}\leq \sqrt{2}.
\end{align*}
Considering this, and taking into account that $0\leq \left\langle\Pi_{{\rm odd}\, i; i\not\in S}A^+_{x_{i}}\Pi_{{\rm even}\, j}A_{x_{j}=0}\right\rangle^{\frac{2}{n-1}}\leq 1$ and that $0\leq \left\langle\Pi_{{\rm odd}\, i; i\not\in S }A^-_{x_{i}}\Pi_{{\rm even}\, j}A_{x_{j}=1}\right\rangle^{\frac{2}{n-1}}\leq 1$, it is possible to write that
\begin{equation}
   \begin{aligned}
       &\left\langle\prod\limits_{{\rm odd}\, i; i\not\in S }A^+_{x_{i}}\prod\limits_{{\rm even}\, j}A_{x_{j}=0}\right\rangle^{\frac{2}{n-1}}\leq \sqrt{2}\cos^2\ta,
       \\
       &\left\langle\prod\limits_{{\rm odd}\, i; i\not\in S}A^-_{x_{i}}\prod\limits_{{\rm even}\, j}A_{x_{j}=1}\right\rangle^{\frac{2}{n-1}}\leq \sqrt{2}\sin^2\ta,
   \end{aligned}
   \label{DD3}
\end{equation}
for some $\ta\in [0,\pi/2]$.
Moreover, using the fact that $|\left\langle A_{x_i}\right\rangle|\leq 1$ when $i\in S$ it follows that
\begin{align*}
   & \left|\left\langle \prod\limits_{i\in S}A^{+}_{x_{i}}\pm \prod\limits_{i\in S}A^{-}_{x_{i}}\right\rangle \right|
   \\
    &\qquad\leq\frac{1}{2^{|S|-1}}  \max\left\{\left|\sum_{(x_i,i\in S);\oplus_{i\in S}x_i=0}\left\langle \prod_{i\in S}A_{x_i}\right\rangle \right|\right.,
    \\
    &
    \qquad\qquad\qquad\qquad\quad\left.\left|\sum_{(x_i,i\in S);\oplus_{i\in S}x_i=1}\left\langle\prod_{i\in S}A_{x_i}\right\rangle \right|
    \right\}
    \\
    &\qquad\leq\frac{1}{2^{|S|-1}}  \times 2^{|S|-1} \\
    &\qquad=1,
\end{align*}
because $|\{(x_i,i\in S)|\oplus_{i\in S}x_i=0 \}|=|\{(x_i,i\in S)|\oplus_{i\in S}x_i=1 \}|=2^{|S|-1}$, the  operation $\oplus$ representing addition modulo 2.
It is thus possible to write $\left|\left\langle \prod\limits_{i\in S}A^{+}_{x_{i}}\right\rangle\right|\leq \cos^{2}\vta$ and $\left|\left\langle \prod\limits_{i\in S}A^{-}_{x_{i}}\right\rangle\right|\leq \sin^{2}\vta$ for some $\vta\in [0,\pi/2]$.
Substituting all the above in Eq.~\eqref{DD1} it follows that
\begin{equation}
    \begin{aligned}
        I^{\frac{2}{n+1}}+J^{\frac{2}{n+1}}
        \leq\,&2^{\frac{n-1}{2(n+1)}}\left[\left\langle \prod\limits_{i\in S}A^{+}_{x_{i}}\right\rangle^{\frac{2}{n+1}}\cos^{\frac{2n-2}{n+1}}\ta\right.
        \\
        &\qquad\qquad+\left.\left\langle \prod\limits_{i\in S}A^{-}_{x_{i}}\right\rangle^{\frac{2}{n+1}}\sin^{\frac{2n-2}{n+1}}\ta\right]
        \\
        \leq\,&2^{\frac{n-1}{2(n+1)}}\left(\cos^{\frac{4}{n+1}}\vta \cos^{\frac{2n-2}{n+1}}\ta \right.
        \\
        &    \qquad\qquad+ \left.\sin^{\frac{4}{n+1}}\vta\sin^{\frac{2n-2}{n+1}}\ta\right).
    \end{aligned}
    \label{DD4}
\end{equation}
The unique extreme point of the continuous function $\cos^{\frac{4}{n+1}}\vta \cos^{\frac{2n-2}{n+1}}\ta+\sin^{\frac{4}{n+1}}\vta\sin^{\frac{2n-2}{n+1}}\ta$ is given by $\tan\vta=\tan\ta$.
By defining $\vta:=\ta$, Eq.~\eqref{DD4} reads
\begin{align}
    I^{\frac{2}{n+1}}+J^{\frac{2}{n+1}}
    \leq\, &2^{\frac{n-1}{2(n+1)}}(\cos^2\ta+\sin^2\ta)
    \nonumber\\
    =\,&2^{\frac{n-1}{2(n+1)}}
    \label{DD5}
\end{align}
where the equality holds if and only if $\vta=\ta=\pi/4$.

\textbf{Case} $\mathbf{\ell<(n-1)/2}$:
An example of this is, e.g., when the sources connecting parties $\sA_{2i}$ and $\sA_{2i+1}$ distribute classical variables, and the sources connecting parties $\sA_{2i-1}$ and $\sA_{2i}$ distribute quantum systems (recall that $i=1, \dots, (n+1)/2$).
This network, and in fact any network with fewer than half of its sources being classical, is a $(n+1)/2$-independent network \cite{Luo2018}, and so the joint correlations satisfy the inequality
\begin{eqnarray}
    I^{\frac{2}{n+1}}+J^{\frac{2}{n+1}}\leq \sqrt{2},
    \label{DD6}
\end{eqnarray}
where the equality holds if all the bipartite quantum sources are maximally entangled EPR states \cite{EPR}.

In summary, we have that for chain networks composed of an odd number of parties, $\ell$-QNL correlations satisfy the following inequality:
\begin{equation}
    I^{\frac{2}{n+1}}+J^{\frac{2}{n+1}}
    \leq
    \begin{cases}
        2^{\frac{n-1}{2(n+1)}} & \ell \geq \frac{n-1}{2} \\
        \sqrt{2} & \ell<\frac{n-1}{2}.
    \end{cases}
    \label{nlchain}
\end{equation}
There are a few aspects to note of this inequality.
First, the bounds obtained are valid for any $\ell$ within each of the ranges, and thus, a violation guarantees the strongest case (namely, any value above $2^{\frac{n-1}{2(n+1)}}$ guarantees $\frac{n-1}{2}$-QNN and any value above $\sqrt{2}$ guarantees FQNN).
Second, the bound for $\ell\geq\frac{n-1}{2}$ increases with $n$, which means that it is more demanding to demonstrate $\ell$-QNN with increasing $n$.
Third, the gap between the two bounds closes in the limit of $n\rightarrow\infty$.
Thus, in infinite chain networks demonstrating nonclassicality of even just one of the sources is as demanding as demonstrating nonclassicality of all of them.
Finally, $\sqrt{2}$ is the maximum value achievable with quantum sources \cite{Luo2018}, so Eq.~\eqref{nlchain} cannot be used to witness FQNN.
In Appendix \ref{app:nlchain} we derive analogous inequalities in the case when the chain has an even number of parties, and that are satisfied by $\ell$-NL correlations.

\section{Nonlinear $\ell$-(Q)NN hierarchies for star networks}\label{sec:nlstar}
The bilocality inequality has also been generalized to star networks \cite{TSCA} (recall Fig.~\ref{fig-1}(c)), whereby $n$ branch parties, $\sA_1, \dots, \sA_n$, share each a bipartite system with a central node, $\sB$.
The symmetry of this setup with respect to permutations of the branch parties allows for easier treatment, that leads to families of inequalities detecting $\ell$-(Q)NN for each $\ell$.

\begin{figure}
    \centering
    \includegraphics[width=0.95\columnwidth]{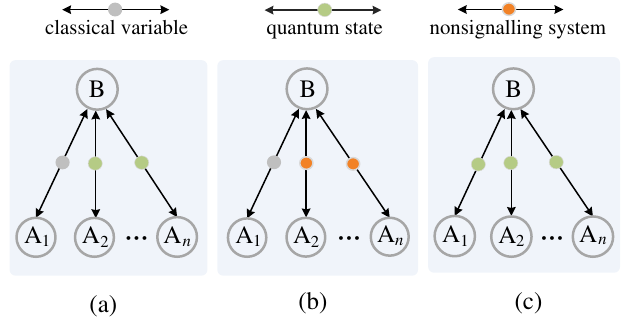}
    \caption{The $n+1$-partite star network. Correlations that do not admit a realization neither in the form of panel (a) nor with any other positioning of the classical variable are FQNN, and those that do not admit a realization neither in the form of panel (b) nor with any other positioning of the classical variable are FNN. There exist correlations of both types that can be produced in the setup depicted in panel (c).
    }
    \label{fig:stars}
\end{figure}

Let us begin with the analysis of $\ell$-QNN for fixed $\ell$.
In this case, we set that the central party can perform two measurements $B_y$, $y\in \{0,1\}$, with binary outcomes. $b\in\{0,1\}$.
In this setup, the quantities $I$ and $J$ are given by $I=\left\langle B_{0}\Pi_{i=1}^nA^+_{x_{i}}\right\rangle$ and $J=\left\langle B_{1}\Pi_{i=1}^nA^-_{x_{i}}\right\rangle$ \cite{TSCA}.
Consider thus the network in Fig.~\ref{fig:stars}(a), where there are $\ell$ parties sharing classical variables with $\sB$ while the remaining ones share quantum states.
Without loss of generality, one can assume that said parties are $\sA_1,\dots,\sA_\ell$.
It follows that
\begin{align}
    |I|^{\frac{1}{n}}+|J|^{\frac{1}{n}}
    =&\left|\left\langle \prod\limits_{i=1}^\ell A^+_{x_{i}}\right\rangle\right|^{\frac{1}{n}}\left|\left\langle B_0\prod\limits_{j=\ell+1}^{n}A^+_{x_{j}}\right\rangle\right|^{\frac{1}{n}}
   \nonumber
   \\
   &+\,\left|\left\langle \prod\limits_{i=1}^\ell A^-_{x_{i}}\right\rangle\right|^{\frac{1}{n}}\left|\left\langle B_1\prod\limits_{j=\ell+1}^{n}A^-_{x_{j}}\right\rangle\right|^{\frac{1}{n}},
    \label{FF1}
\end{align}
where, as in previous sections, $A^\pm_{x_{i}}=(A_{x_i=0}\pm A_{x_{i}=1})/2$.
For any set of quantum observables $A^\pm_{x_{i}}$, the following inequality holds:
\begin{equation*}
\left|\left\langle B_0\prod\limits_{j=\ell+1}^{n}A^+_{x_{j}}\right\rangle\right|^{\frac{1}{n-\ell}}
+\left|\left\langle B_1\prod\limits_{j=\ell+1}^{n}A^-_{x_{j}}\right\rangle\right|^{\frac{1}{n-\ell}}\leq \sqrt{2}.
\end{equation*}
This follows from the quantum bound of the nonlinear Bell-type inequality for the $n-k$-independent quantum network in Ref.~\cite{Luo2018}.
Therefore, one can write that
\begin{equation}
    \begin{aligned}
        &\left|\left\langle B_0\prod\limits_{j=\ell+1}^{n}A^+_{x_{j}}\right\rangle\right|^{\frac{1}{n-\ell}}\leq \sqrt{2}\cos^2\ta, 
        \\
        &\left|\left\langle B_1\prod\limits_{j=\ell+1}^{n}A^-_{x_{j}}\right\rangle\right|^{\frac{1}{n-\ell}}\leq \sqrt{2}\sin^2\ta,
    \end{aligned}
    \label{qstarbounds}
\end{equation}
for some $\ta\in [0,\pi/2]$.
Additionally, since all the parties $\sA_1, \dots, \sA_\ell$ are independent, we have the classical bounds $\left|\left\langle \prod_{i=1}^\ell A^\pm_{x_i}\right\rangle\right|\leq \prod_{i=1}^\ell \left|\left\langle A^\pm_{x_i}\right\rangle\right|$.
In combination with $\left|\left\langle A^{+}_{x_{i}}\pm A^{-}_{x_{i}}\right\rangle\right|=\left|\left\langle A_{x_i}\right\rangle\right|\leq 1$, one can define $\left|\left\langle \prod_{i=1}^\ell A^+_{x_i}\right\rangle\right|\leq \prod_{i=1}^\ell\cos^{2}\vta_i$ and $\left|\left\langle \prod_{i=1}^\ell A^-_{x_i}\right\rangle\right|\leq \prod_{i=1}^\ell\sin^{2}\vta_i$ for some $\vta_i\in [0,\pi/2]$.
Substituting this in Eq.~\eqref{FF1} we have that
\begin{align*}
    |I|^{\frac{1}{n}}+|J|^{\frac{1}{n}}
    =2^{\frac{n-\ell}{2n}}&\left(\prod_{i=1}^\ell\cos^{\frac{2}{n}}\vta_i \cos^{\frac{2n-2\ell}{n}}\ta\right.
    \\
    &\quad\left.+\prod_{i=1}^\ell\sin^{\frac{2}{n}}\vta_i\sin^{\frac{2n-2\ell}{n}}\ta\right).
\end{align*}
The maximum of the right-hand side is achieved at $\vta_1=\dots =\vta_\ell:=\vta$.
This implies that 
\begin{equation*}
    |I|^{\frac{1}{n}}+|J|^{\frac{1}{n}}
    \!\leq\! 2^{\frac{n-\ell}{2n}}\!\!\left(\cos^{\frac{2\ell}{n}}\vta \cos^{\frac{2n-2\ell}{n}}\ta \!+\!\sin^{\frac{2\ell}{n}}\vta\sin^{\frac{2n-2\ell}{n}}\ta\right)\!.
\end{equation*}
Finally, the continuous function $\cos^{\frac{2\ell}{n}}\vta \cos^{\frac{2n-2\ell}{n}}\ta+\sin^{\frac{2\ell}{n}}\vta\sin^{\frac{2n-2\ell}{n}}\ta$ has a unique extreme point, given by $\vta=\ta$.
We thus have that
\begin{align}
    |I|^{\frac{1}{n}}+|J|^{\frac{1}{n}}
    \leq\, &2^{\frac{n-\ell}{2n}}
    \left(\cos^2\ta+\sin^2\ta\right)
    \nonumber\\
    =\,&2^{\frac{n-\ell}{2n}}.
    \label{FF5}
\end{align}

The right-hand side of Eq.~\eqref{FF5} has a different value for each $\ell$, thereby giving a hierarchy of inequalities that detects $\ell$-QNN for every $\ell\in[1,\cdots,m]$.

Consider now the situation when there are $\ell$ extremal parties (e.g., $\sA_1,\dots,\sA_\ell$) that share classical randomness with $\sB$ while the remaining parties share nonsignaling systems.
In this case, the bounds in Eq.~\eqref{qstarbounds} are replaced by $\left|\left\langle B_{0}\Pi_{i>\ell}A^+_{x_{i}}\right\rangle\right|\leq 1$, $\left|\left\langle B_1\Pi_{i>\ell}A^-_{x_{i}}\right\rangle\right|\leq 1$, since there exist distributions built out of PR boxes that saturate these bounds, in analogy with the construction in Appendix \ref{app:lnnchain} for chain networks.
We thus have that
\begin{align}
    |I|^{\frac{1}{n}}+|J|^{\frac{1}{n}}\leq\, & \left|\left\langle\prod\limits_{i=1}^{\ell}A^+_{x_{i}}\right\rangle\right|^{\frac{1}{n}}+
    \left|\left\langle\prod\limits_{i=1}^{\ell}A^-_{x_{i}}\right\rangle\right|^{\frac{1}{n}}
    \nonumber
    \\
    \leq\, &\prod\limits_{i=1}^{\ell}\cos^{\frac{2}{n}}\vta_i +\prod_{i=1}^{\ell}\sin^{\frac{2}{n}}\vta_i
    \label{F7}
    \\
    :=&\cos^{\frac{2\ell}{n}}\vta +\sin^{\frac{2\ell}{n}}\vta
    \label{F8}
    \\
    \leq\,&  2^{\frac{n-\ell}{n}},
    \label{F9}
\end{align}
where we have used the fact that $\left|\left\langle A^+_{x_{i}}\right\rangle\right|\leq \cos^2\vta_i$ and $\left|\left\langle A^-_{x_{i}}\right\rangle\right| \leq \sin^2\vta_i$ for some $\vta_i$ in Eq.~\eqref{F7}, which stem from the inequality $\left|\left\langle A^+_{x_{i}}\pm A^-_{x_{i}}\right\rangle\right|\leq 1$.
Inequality \eqref{F8} follows in a way analogous to that of $\ell$-QNN studied above, and
Eq.~\eqref{F9} follows from computing the unique extreme point of the continuous function $\cos^{\frac{2\ell}{n}}\vta +\sin^{\frac{2\ell}{n}}\vta$, which is $\vta=\pi/4$.
We thus, again, obtain a hierarchy of bounds that detects $\ell$-NN for every $\ell$.
Note also that, for any placement of the classical sources, the same inequality is obtained, and thus a violation of the single inequality \eqref{F9} is a certification of $\ell$-NN.

Both sets of inequalities, Eqs.~\eqref{FF5} and \eqref{F9}, can be violated in quantum mechanics.
If each of the sources distributes a bipartite quantum system in the state given by Eq.~\eqref{A006}, and the parties perform the measurements given by $A_{x_i} =\cos\vta \sigma_3+(-1)^{x_i}\sin\vta\sigma_1$ and $B_y=(1-y)\sigma_3^{\otimes n}+y\sigma_1^{\otimes n}$ \cite{Luo2018} one has that 
\begin{equation*}
    |I|^{\frac{1}{n}}+|J|^{\frac{1}{n}}=
    \cos\vta+\sin\vta\prod\limits_{i=1}^n\sin^{\frac{1}{n}}2\ta_i.
\end{equation*}
If we choose $\vta$ such that $\cos\vta=(1+\prod_{i=1}^n\sin^{\frac{2}{n}}2\ta_i)^{-\frac12}$, we have a violation of Eq.~\eqref{FF5} whenever $\prod_{i=1}^n\sin^{\frac{2}{n}}2\ta_i>2^{\frac{n-\ell}{n}}-1$, and a violation of Eq.~\eqref{F9} when $\prod_{i=1}^n\sin^{\frac{2}{n}}2\ta_i>2^{\frac{2n-2\ell}{n}}-1$.
Take as an example the case when the sources distribute maximally entangled EPR states, i.e., $\theta_i=\pi/4$ $\forall\,i$.
In this case the value achieved is $|I|^{\frac{1}{n}}+|J|^{\frac{1}{n}}=\sqrt{2}$.
This is larger than $2^{\frac{n-\ell}{n}}$ for all $\ell>n/2$, so Eq.~\eqref{F9} is useful for verifying $\ell$-NN in such star networks only for $\ell>n/2$.

\section{Linear FNN inequalities for chains}
Up to now, we have used nonlinear inequalities to detect $\ell$-NN and $\ell$-QNN.
On one hand, nonlinear inequalities are useful because they approximate better the boundaries of the relevant sets, which are known to be nonlinear \cite{BRGP}. However, the nonlinearities make it difficult to optimize these inequalities and to find explicit models.
In this section and in the following one we present linear inequalities whose violations detect FNN.
For deriving them, we exploit the fact that realizing entanglement swapping is not possible when parties are connected by classical sources.
It is known that at least some supra-quantum theories do not allow for entanglement swapping \cite{Short2006}, and this fact has been exploited to develop NN inequalities in the tripartite chain scenario \cite{Weilenmann2020}.
However, we are interested in guaranteeing nonclassical phenomena, so as long as there exists a nonclassical theory that allows for entanglement swapping (as it is the case of quantum mechanics), the inequalities below can, in principle, be used for detecting FNN.

\subsection{The tripartite chain}
Let us begin, as in Section~\ref{chainna}, with analyzing the tripartite chain scenario, depicted in Fig.~\ref{fig-1} (a).
A way to guarantee that both sources are not classical is by completing entanglement swapping, so that after Bob's measurement the joint state of Alice and Charlie can violate the CHSH inequality.
Using this fact, consider the following operator, where Alice can perform two dichotomic measurements, Bob performs a single, four-outcome measurement, and Charlie can perform three dichotomic measurements:
\begin{equation*}
    \begin{aligned}
        \cB_3:=\,&A_0B^{0}C_0+A_0B^{1}C_0+A_0B^{0}C_1-A_0B^{1}C_1\\
            &+A_1B^{0}C_0+A_1B^{1}C_0-A_1B^{0}C_1+A_1B^{1}C_1\\
            &+A_0B^{2}C_1-A_0B^{3}C_1+A_0B^{2}C_2+A_0B^{3}C_2\\
            &-A_1B^{2}C_1+A_1B^{3}C_1+A_1B^{2}C_2+A_1B^{3}C_2.
    \end{aligned}
\end{equation*}

In the case where Bob and Charlie share classical randomness, the operator $\cB_3$ is a linear combination of Charlie's operators, and since Charlie only receives classical shared randomness we can, without loss of generality, assume deterministic outputs and have $C_i\in \{\pm 1\}$.
By substituting all possibilities for $C_i$, one arrives to
\begin{align*}
    \left\langle \cB_3\right\rangle=\,&\left\langle C_0(B^{0}A_0+B^{0}A_1+B^{1}A_0+B^{1}A_1)\right\rangle \\
    &+\left\langle C_1(B^{0}A_0-B^{0}A_1-B^{1}A_0+B^{1}A_1)\right\rangle \\
    &+\left\langle C_1(B^{2}A_0-B^{2}A_1-B^{3}A_0+B^{3}A_1)\right\rangle \\
    &+\left\langle C_2(B^{2}A_0+B^{2}A_1+B^{3}A_0+B^{3}A_1)\right\rangle \\
    \leq\, & \max\{\pm \langle 2(B^{0}+B^{2})A_0+2(B^{1}+B^{3})A_1\rangle,
   \\
    & \pm \langle 2(B^{0}+B^{2})A_1+2(B^{1}+B^{3})A_0\rangle, \\
    &\pm \langle 2(B^{0}-B^{3})A_1+2(B^{1}-B^{2})A_0\rangle,
  \\
    &  \pm \langle 2(B^{3}-B^{0})A_0+2(B^{2}-B^{1})A_1\rangle\} \\
    \leq\, &2,
\end{align*}
where the last bound can be found by individually obtaining the maximum value for each case, using $|\left\langle A_i\right\rangle|\leq 1$ and $\sum_{i=0}^3 B^i=\openone$.

Similarly, consider the alternative scenario, where Alice and Bob share the classical source.
Since $\cB_3$ is also linear in Alice's measurements, one can proceed analogously by assuming deterministic outputs for Alice, so $A_i\in \{\pm 1\}$, arriving to
\begin{align*}
    \left\langle \cB_3\right\rangle=\,&\left\langle A_0(B^{0}C_0+B^{1}C_0+B^{0}C_1-B^{1}C_1)\right\rangle \\
    &+\left\langle A_1(B^{0}C_0+B^{1}C_0-B^{0}C_1+B^{1}C_1)\right\rangle \\
    &+\left\langle A_0(B^{2}C_1-B^{3}C_1+B^{2}C_2+B^{3}C_2)\right\rangle \\
    &+\left\langle A_1(-B^{2}C_1+B^{3}C_1+B^{2}C_2+B^{3}C_2)\right\rangle \\
    \leq\, & \max\left\{\pm \left\langle 2(B^{0}+B^{1})C_0+2(B^{2}+B^{3})C_2\right\rangle,\right.\\
    &\qquad\,\,\,\left.\pm \left\langle 2(B^{0}-B^{1})C_1+2(B^{2}-B^{3})C_2\right\rangle\right\} \\
    \leq\, &2.
\end{align*}

This means that the inequality
\begin{equation}
    \begin{aligned}
        &\left\langle A_0B^{0}C_0+A_0B^{1}C_0+A_0B^{0}C_1-A_0B^{1}C_1\right\rangle\\
        &+\left\langle A_1B^{0}C_0+A_1B^{1}C_0-A_1B^{0}C_1+A_1B^{1}C_1\right\rangle\\
        &+\left\langle A_0B^{2}C_1-A_0B^{3}C_1+A_0B^{2}C_2+A_0B^{3}C_2\right\rangle\\
        &-\left\langle A_1B^{2}C_1+A_1B^{3}C_1+A_1B^{2}C_2+A_1B^{3}C_2\right\rangle
        \leq 2
    \end{aligned}
    \label{lbilocal}
\end{equation}
holds whenever one of the sources distributes classical systems, regardless of the nature of the other and of which specific source is the one distributing the classical systems.
Its maximal quantum value is $2\sqrt{2}$, which can be derived directly from the upper bound to the CHSH inequality.
Moreover, quantum strategies based on maximally entangled states saturate this bound (see Appendix~\ref{app:lchain} for an example realization).

\subsection{Long chains}
The reasoning above can be generalized to chain networks of arbitrary length, by exploiting the fact that nonclassical correlations between the extremal parties can only be established if all the sources are nonclassical.
Consider thus the following operator:
\begin{equation*}
    \begin{aligned}
      \cB_n:=&A_0^{(1)}B^{00}A_0^{(n)}+A_0^{(1)}B^{01}A_0^{(n)}+A_0^{(1)}B^{00}A_1^{(n)}\\
        &-A_0^{(1)}B^{01}A_1^{(n)}+A_1^{(1)}B^{00}A_0^{(n)}+A_1^{(1)}B^{01}A_0^{(n)}\\
        &-A_1^{(1)}B^{00}A_1^{(n)}+A_1^{(1)}B^{01}A_1^{(n)}+A_0^{(1)}B^{10}A_1^{(n)}
        \\
        &-A_0^{(1)}B^{11}A_1^{(n)}+A_0^{(1)}B^{10}A_2^{(n)}+A_0^{(1)}B^{11}A_2^{(n)}\\
        &-A_1^{(1)}B^{10}A_1^{(n)}+A_1^{(1)}B^{11}A_1^{(n)}+A_1^{(1)}B^{10}A_2^{(n)}
        \\
        &+A_1^{(1)}B^{11}A_2^{(n)},
    \end{aligned}
\end{equation*}
where the extremal parties are $\sA_1$ and $\sA_n$, which have operators $A^{(1)}_i$ and $A^{(n)}_i$, respectively, and $B^{st}$ are products of operators of all the non-extremal parties $\sA_2,\cdots, \sA_{n-1}$, given by $B^{st}=\sum_{^{\oplus_{k=2}^{n-1} i_k j_k=st}}M^{i_2j_2}\cdots M^{i_{n-1}j_{n-1}}$, with $\oplus$ denoting the bitwise XOR operation, i.e., $xy \oplus uv=(x \oplus u, y\oplus v)$, and $M^{i_k j_k}$ are the projectors corresponding to the single measurement of party $\sA_k$. Similarly to Eq.~\eqref{lbilocal}, the extremal party $\sA_n$ has three dichotomic measurements.

Let us begin by assuming that the classical source is that connecting $\sA_1$ and $\sA_2$.
This network can be understood as a tripartite chain network where the central party is composed of parties $\sA_2,\dots,\sA_{n-1}$. Then, following the previous section, we can assume deterministic outputs for $A_{x_1}^{(1)}\in \{\pm 1\}$, finding that
\begin{align*}
    &\left\langle \cB_n \right\rangle
    \\
    &\quad=\left\langle A_0^{(1)}(B^{00}A_0^{(n)}\!+\!B^{01}A_0^{(n)}\!+\!B^{00}A_1^{(n)}\!-\!B^{01}A_1^{(n)})\right\rangle \\
    &\qquad+\left\langle A_1^{(1)}(B^{00}A_0^{(n)}\!+\!B^{01}A_0^{(n)}\!-\!B^{00}A_1^{(n)}\!+\!B^{01}A_1^{(n)})\right\rangle \\
    &\qquad+\left\langle A_0^{(1)}(B^{10}A_1^{(n)}\!-\!B^{11}A_1^{(n)}\!+\!B^{10}A_2^{(n)}\!+\!B^{11}A_2^{(n)})\right\rangle \\
    &\qquad+\left\langle A_1^{(1)}(\!-\!B^{10}A_1^{(n)}\!+\!B^{11}A_1^{(n)}\!+\!B^{10}A_2^{(n)}\!+\!B^{11}A_2^{(n)})\right\rangle \\
    &\quad \leq\, \max\left\{\pm\! \left\langle 2(B^{00}+B^{01})A_0^{(n)}+2(B^{10}+B^{11})A_2^{(n)}\right\rangle\!,\right.
    \\
    &
    \qquad\qquad\quad\pm\!\! \left.\left\langle 2(B^{00}-B^{01})A_1^{(n)}+2(B^{10}-B^{11})A_2^{(n)}\right\rangle\right\} \\
    &\quad\leq\, 2,
\end{align*}
where the terms in the maximization are found by applying the triangle inequality and forming groups that can be jointly maximized.
Then, the bound follows from optimizing individually each of the elements under the constraints $\sum_{b_0,b_1}B^{b_0b_1}=\openone$ and $\left|\left\langle A_i^{(n)}\right\rangle\right|\leq 1$.

Now, consider the case where the classical source is not in the extreme of the chain.
In that case, we can rewrite $\cB_n$ as
\begin{align*}
        &\cB_n=
        \\
        &\quad B^{00}\left(A_{0}^{(1)}A_{0}^{(n)} + A_{0}^{(1)}A_{1}^{(n)} + A_{1}^{(1)}A_{0}^{(n)} - A_{1}^{(1)}A_{1}^{(n)}\right)\\
        &\qquad+B^{01}\left(A_{0}^{(1)}A_{0}^{(n)}\!-\!A_{0}^{(1)}A_{1}^{(n)}\!+\!A_{1}^{(1)}A_{0}^{(n)}\!+\!A_{1}^{(1)}A_{1}^{(n)}\right)\\
        &\qquad+B^{10}\left(A_{0}^{(1)}A_{1}^{(n)}\!-\!A_{0}^{(1)}A_{2}^{(n)}\!+\!A_{1}^{(1)}A_{1}^{(n)}\!+\!A_{1}^{(1)}A_{2}^{(n)}\right)\\
        &\qquad+B^{11}\left(\!-\!A_{0}^{(1)}A_{1}^{(n)}\!+\!A_{0}^{(1)}A_{2}^{(n)}\!+\!A_{1}^{(1)}A_{1}^{(n)}\!+\!A_{1}^{(1)}A_{2}^{(n)}\right)\!.
\end{align*}

Each line above corresponds to a different CHSH test, weighted by a different probability of obtaining certain outcomes in the central parties (recall that the $B^{st}$ are defined by $B^{st}=\sum_{\oplus_{\ell=2}^{n-1} i_\ell{}j_\ell=st}M^{i_2j_2}\dots M^{i_{n-1}j_{n-1}}$, with $\oplus$ being the bit-wise XOR).
Therefore, the expectation value of each of the lines above can be bounded by $2\sum_{\oplus_{i=2}^{n-1}a_i=00} p(a_2,\dots,a_{n-1})$, and using the fact that $\sum_{i=2}^{n-1}\sum_{a_i} p(a_2,\dots,a_{n-1})=1$ the upper bound of $2$ follows.
We are thus left with the inequality
\begin{equation}
    \begin{aligned}
        \left\langle \cB_n\right\rangle
        &=\left\langle A_0^{(1)}\left[(B^{00}+B^{01})A_0^{(n)}+(B^{00}-B^{01})A_1^{(n)}\right]\right\rangle \\
        &\quad+\left\langle A_1^{(1)}\left[(B^{00}+B^{01})A_0^{(n)}+(B^{00}-B^{01})A_1^{(n)}\right]\right\rangle \\
        &\quad+\left\langle A_0^{(1)}\left[(B^{10}-B^{11})A_1^{(n)}+(B^{10}+B^{11})A_2^{(n)}\right]\right\rangle \\
        &\quad+\left\langle A_1^{(1)}\left[(-B^{10}+B^{11})A_1^{(n)}\!+\!(B^{10}+B^{11})A_2^{(n)}\right]\right\rangle \\
       & \leq 2.
    \end{aligned}
    \label{chainl}
\end{equation}

The maximal quantum bound for Eq.~\eqref{chainl} is, as before, $2\sqrt{2}$ from the CHSH inequality.
As in the previous cases, this is a single inequality whose violation certifies that all sources in the network are not classical, i.e., it detects FNN in chain networks.
Moreover, it can be violated within quantum mechanics, as we also show in Appendix~\ref{app:lchain}.

\section{Linear FNN inequalities for stars}
It is possible to use the results of the previous section to witness FNN in star networks, in two different ways.
The first one is by decomposing the star network into a collection of tripartite chain subnetworks comprised by the parties $(\sA_{i},\sA_{j\not=i},\sB)$, and testing Eq.~\eqref{lbilocal} on each of them.
This leads to witnesses in scenarios where the central party performs a single, four-outcome measurement, and the branch parties can perform one of either two or three measurements, depending on the particular inequalities that are used for the testing.

The second one is by exploiting the same principle, namely that only measurements on nonclassical systems can produce nonlocal correlations.
In this case, we are interested in detecting genuinely multipartite nonlocality in the branch parties after the measurement of the central party.
One way of detecting genuine multipartite nonlocality for $n$ parties is via the Mermin-Svetlichny inequality \cite{Svetlichny,Mermin,Collins}:
\begin{equation}
    \left\langle \mathcal{S}\left(A_{x_1}^{(\vi)},\dots, A_{x_n}^{(\vi)}\right)\right\rangle\leq 2^{n-1},
    \label{eq:svet}
\end{equation}
where the operator $\cS\left(A_1,\dots, A_n\right)\equiv S_n$ is built recursively via
\begin{equation*}
    S_n=(A_{x_n=0}+A_{x_n=1})S_{n-1}+(A_{x_n=0}-A_{x_n=1})S_{n-1}',
\end{equation*}
with $S_1=A_{x_1=0}$, and $S_k'$ being obtained from $S_k$ by flipping all the inputs of all the parties.
By creating one such operator for every outcome of the central party $\sB$, one obtains the inequality
\begin{eqnarray}
    \left\langle \sum_{\vi}M^{\vi}\cS^{(\vi)}\left(A_{x_1}^{(\vi)}, \dots,A_{x_n}^{(\vi)}\right)\right\rangle\leq 2^{n-1},
    \label{starl}
\end{eqnarray}
where $\{M^{\vi}\}$ denotes the projection measurement of $\sB$ with outcome $\vi:=i_1\dots i_n$, and $A_{x_j}^{_{(\vi)}}$ represents the corresponding measurement for party $\sA_j$.
If one of the sources in the network is classical, then the measurement of the central party projects the joint state of the extremal parties into a biseparable-type state, thus satisfying Eq.~\eqref{eq:svet} for any measurements $A_{x_1}^{(\vi)},\dots, A_{x_n}^{(\vi)}$ \cite{Svetlichny,Mermin,Collins}, and thus satisfying Eq.~\eqref{starl}.

Equivalently, from the quantum bound of the Mermin-Svetlichny operator it is easy to see that the maximum quantum value of Eq.~\eqref{starl} is $2^{n-1}\sqrt{2}$.
A violation of Eq.~\eqref{starl} can be achieved, for instance, having the sources distribute states of the form in Eq.~\eqref{A006}.
Take the measurement of $\sB$ to be given by $\{M^{\vi}\}$ with
\begin{equation*}
    M^{\vi}=\ketbra{\phi_{0i_{[n-1]}}^{i_1}}{\phi_{0i_{[n-1]}}^{i_1}},
\end{equation*}
where $i_{[n-1]}:=i_2\dots i_n$, $\ket{\phi_{0i_{[n-1]}}^{i_1}}=\frac{1}{\sqrt{2}}(\ket{0,i_{[n-1]}} + (-1)^{i_1}\ket{1,j_{[n-1]}})$, and $j_{[n-1]}$ denotes the complement bit series of $i_{[n-1]}$, i.e., $i_{[n-1]}\oplus j_{[n-1]}=1\dots 1$ with $\oplus$ being the bitwise XOR.
For each measurement outcome $\vi$, with probability $p(\vi)$, the joint system of particles owned by all $\sA_j$s will collapse into
\begin{equation*}
    \ket{\Phi_{0i_{[n-1]}}}=\gamma_{0i_{[n-1]}}\ket{0,i_{[n-1]}} +(-1)^{i_1}\delta_{1i_{[n-1]}}\ket{1,j_{[n-1]}},
\end{equation*}
where $\gamma_{0i_{[n-1]}}$ and $\delta_{1i_{[n-1]}}$ are defined by
\begin{equation*}
    \gamma_{0i_{[n-1]}}=\frac{1}{\sqrt{p(\vi)}}\alpha_0\prod_{j=2}^n\alpha_{i_j},
    \quad
    \delta_{1i_{[n-1]}}=\frac{1}{\sqrt{p(\vi)}}\alpha_1\prod_{j=2}^n\alpha_{\overline{i}_j},
\end{equation*}
with $p(\vi)=\alpha^2_0\prod_{j=2}^n\alpha_{i_j}^2
+\alpha_1^2\prod_{j=2}^n\alpha_{\overline{i}_j}^2$, $\alpha_{i_j=0}=\cos\ta_j$ and $\alpha_{i_j=1}=\sin\ta_j$, $j=1, \dots, n$.
This type of state can be transformed into generalized GHZ states \cite{GHZ} by using local unitary operations $U_{i_2},\dots, U_{i_n}$, i.e., $\ket{\text{GHZ}_{0i_{[n-1]}}}:=\otimes_{j=1}^n U_{i_j}\ket{\Phi_{0i_{[n-1]}}}$.

Now, define the quantum observables of $\sA_{j}$ conditional on the outcome $\vi$ as $A_{a_j=i_j}^{(i_{[n-1]})}\in \{U_{i_j}^\dag(\cos\vta_{\vi}\sigma_1
\pm \sin\vta_{\vi}\sigma_y)U_{i_j}\}$, with $\vta_{\vi}\in (0,\pi)$ and $j=1, \dots, n$.
From Ref.~\cite{Zukowski2002}, we have that by choosing proper phases $\vta_{\vi}$ one can obtain that $\bra{\text{GHZ}_{0i_{[n-1]}}}\mathcal{S}(A_{1}^{(\vi)},\dots, A_n^{(\vi)})\ket{\text{GHZ}_{0i_{[n-1]}}}$ exceeds $2^{n-1}$ for some $\ta_j$ satisfying $2\gamma_{0i_{[n-1]}}\delta_{1i_{[n-1]}}>1/\sqrt{2^{n-1}}$.
Therefore, using the necessary phases $\vta_{\vi}$, we have a violation of Eq.~\eqref{starl} whenever the $\ta_j$ satisfy the condition $\min_{i_{[n-1]}}\{2\gamma_{0i_{[n-1]}}\delta_{0i_{[n-1]}}\}
>1/\sqrt{2^{n-1}}$.

\section{Inequalities for general networks}
\begin{figure}
    \centering
    \includegraphics[width=0.95\columnwidth]{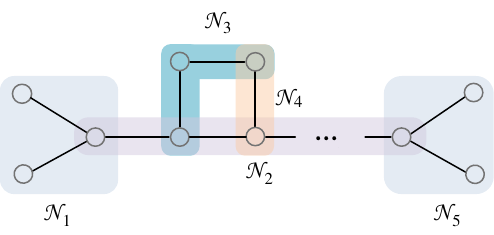}
    \caption{\small Decomposition of an arbitrary network into stars and chains. Here, the network can be decomposed into two star subnetworks ($\cN_1$ and $\cN_5$) and three chain subnetworks ($\cN_2$, $\cN_3$ and $\cN_4$).}
    \label{fig:decompo}
\end{figure}

In general, one can use all the inequalities derived in this work to test for FQNN and FNN in arbitrary networks.
For the case of networks composed exclusively of bipartite sources, one can decompose the network in a set of chain and star networks (see Fig.~\ref{fig:decompo} for an example), and test F(Q)NN in the different subnetworks using Eqs.~\eqref{nlchain}, \eqref{FF5}, \eqref{F9}, \eqref{chainl}, and \eqref{starl}.
In particular, cycles can be decomposed in terms of chain networks.
Since all sources belong to at least one subnetwork, observation of F(Q)NN in all subnetworks implies that the full correlation is F(Q)NN.

When the networks contain sources that distribute systems to more than two parties, 
one can use the strategy that we used for star networks, namely building a Svetlichny-type inequality of the form of Eq.~\eqref{starl} for all the parties that share states with parties that receive from the multipartite source, considering as the operators $M$ the projections on all possible outcomes of the parties that are connected by the multipartite source.
In this case, if there are $k$ parties that share states with parties that receive from the multipartite source, non-FNN correlations satisfy
\begin{equation}
    \left\langle \sum_{\vt}M^{\vt}\cS^{(\vt)}\left(B_{x_{i_1}}^{(\vt)},\dots,B_{x_{i_k}}^{(\vt)}\right)\right\rangle\leq 2^{k-1}.
    \label{starll}
\end{equation}

\section{Noise robustness}\label{app:exp}
To conclude, we present a few remarks on the robustness to noise of the inequalities developed in this work.
In order to analyze the robustness to experimental imperfections of the inequalities in the manuscript, we consider quantum realizations based on Werner states, $\rho_{v_i}$, given by
\begin{equation*}
    \rho_{v_i}=v_i\ket{\phi_i}\langle{\phi_i}|+\frac{1-v_i}{d^2}I_{d^2},
\end{equation*}
where $\ket{\phi_i}$ are arbitrary bipartite entangled pure states with local dimension $d$, $I_{d}$ is the $d\times d$ identity matrix, and $0\leq v_i\leq 1$, $i=1, \dots, m$.

Using the states $\ket{\phi_i}$ and the measurements that achieve the quantum violations described throughout the manuscript (see, e.g., Section \ref{sec:nlstar} and Appendix \ref{app:nlchain:quantum}), the corresponding quantities $I$ and $J$ get multiplied by a factor $\prod_{i=1}^m v_i$.
This implies that, in order to observe a violation, the visibilities of the sources must satisfy $\prod_{i=1}^mv_i>\frac{1}{\sqrt{2}}$ when using maximally entangled states.
Similarly, using maximally entangled states for $\ket{\phi_i}$ in the $n+1$-partite star network, $\ell$-NN correlations can be detected via the violation of Eq.~\eqref{F9} for $\prod_{i=1}^mv_i>2^{-\left(\ell-\frac{n}{2}\right)}$, while $\ell$-QNN correlations can be detected via the violation of Eq.~\eqref{FF5} whenever $\prod_{i=1}^m v_i>2^{-\frac{\ell}{2}}$.
As discussed in Section~\ref{sec:nlstar}, this shows that Eq.~\eqref{F9} cannot be used for detecting FNN (although it can be used for detecting $\ell$-NN as long as $\ell>n/2$), while Eq.~\eqref{FF5} can be used for detecting $\ell$-QNN in a noise-robust manner independently of the number of sources in the network, since the critical visibility only depends on $\ell$.

For the case of the linear inequalities for detecting FNN, both allow for violations when using Werner states as long as $\prod_{i=1}^mv_i>1/\sqrt{2}$.
When instead using noisy generalized EPR states \eqref{A006}, the critical visibilities needed for observing a violation of Eq.~\eqref{chainl} relate to the angles $\theta_i$ via
\begin{equation*}
    \prod_{i=1}^nv_i>\frac{1}{\sqrt{1+\prod_{j=1}^{n-1}\sin^{2}2\ta_j}}.
\end{equation*}
Analogously, when decomposing star networks into sets of tripartite chain networks, FNN can in principle be guaranteed when every pair of visibilities satisfies
\begin{equation*}
    v_i v_j > \frac{1}{\sqrt{1+\sin^{2}2\ta_i\sin^{2}2\ta_j}}.
\end{equation*}
Note how, in both cases, the lowest requirements on the visibilities are obtained for $\ta_i=\pi/4$ $\forall\,i$, i.e., for maximally entangled states.

\section{Discussion}
In this work we have provided Bell-like inequalities that witness genuinely quantum phenomena in networks, in the sense that their violations by particular distributions guarantee that these cannot be reproduced when at least a number of the sources in the network are classical.
We have done so using two different assumptions on the nonclassical sources.
First we assumed that they are only constrained by the no-signaling principle.
This leads to a hierarchy of levels of network nonlocality, that interpolates from the standard definition of network nonlocality \cite{networkReview} to that of full network nonlocality \cite{Alej2022}.
Second, we assumed that the sources distribute systems described by quantum mechanics.
This gives rise to a hierarchy analogous to the previous one, that has milder requirements for experimental observations and is thus more suitable for applications in quantum communication. More concretely, we have provided Bell-like inequalities whose violation guarantees (i) $n/2$-QNN and $\ell$-NN in the $n$-partite chain network (Eq.~\eqref{nlchain}, and Eq.~\eqref{DD15} in Appendix \ref{app:nlchain}, respectively), (ii) $\ell$-NN and $\ell$-QNN for any $\ell$ in the $n+1$-partite star network (Eqs. \eqref{F9} and \eqref{FF5}, respectively), and (iii) FNN in arbitrary chain and star networks (Eqs.~\eqref{chainl} and \eqref{starl}, respectively).

Being able to guarantee, from observed statistics, the absence of classical sources in a network is an important step in the development of network-based quantum information protocols.
Concrete protocols exist, for instance for quantum conference key agreement \cite{Hahn2020,Das2021,Pickston2023} or quantum secret sharing \cite{Moreno2020,Luo2022}. However, their proofs of security rely on guaranteeing global multipartite nonclassicality, which leads to demanding requirements. Our work paves the way to the development of protocols and security proofs that exploit the network structure, which has received limited attention so far \cite{Lee2018,Qinflation,Luo2022}.
Indeed, consider the $n+1$-partite star network. Following Refs.~\cite{Lee2018,Luo2022} one can bound the predictive power of an adversary for all outcomes $\va$ by the value obtained in Eq.~\eqref{FF5} via
\begin{equation*}
    \begin{aligned}
        D(P(\ve|\va;\vx,y,\vz),\prod_{i=1}^nP(e_i|z_i)) 
        \leq (2-\mathcal{S}_\ell)n,
    \end{aligned}
\end{equation*}
where $D(p,q)$ is the total variation distance between distributions $p$ and $q$, and $\mathcal{S}_\ell$ is the value achieved for Eq.~\eqref{FF5}.
By substituting the bounds for $\mathcal{S}_\ell$ in Eq.~\eqref{FF5}, it is possible to bound the predictive power of an eavesdropper if the existence of $\ell$ classical sources in the network cannot be discarded.
This, and more importantly, the demonstrations of quantum realizations that lead to FNN and FQNN that we show in this work, strongly motivate the development of novel quantum information protocols tailored to networks. 

Having inequalities for detecting $\ell$-NN and $\ell$-QNN allows to easily see how inserting the assumption that nature is governed by quantum mechanics results in milder conditions to guarantee nonclassicality, which is relevant for experimental realizations.
For instance, detection of FNN ($\ell=1$) in the star network with $n=3$ via Eq.~\eqref{F9} is not possible with noisy versions of the generalized EPR states considered throughout for any visibility, while detection of FQNN is possible when the product of the states' visibilities exceeds $2^{-1/6}\approx0.891$.

All the inequalities we have presented require some of the parties to have a choice of input. However, a fundamental aspect of network nonlocality is that it can be realized in setups without input choices \cite{Fritz,MRen2019,Abiuso2021,Boreiri2023,Pozas2023}. Seeing whether the hierarchies of nonlocality developed here are also present in scenarios without inputs is an interesting problem that we leave open, and that can have consequences useful in applications such as quantum random number generation.
Numerical tools allow to easily address these scenarios for arbitrary networks (see, e.g., \cite{Boghiu2023} or the computational appendix of \cite{Alej2022}), thus in principle permitting the exploration of $\ell$-(Q)NN beyond the situations discussed here.

\begin{acknowledgments}
This work was supported by the National Natural Science Foundation of China (Nos. 62172341, 61772437), Sichuan Natural Science Foundation (No. 2023NSFSC0447), Interdisciplinary Research of Southwest Jiaotong University China (No.2682022KJ004), the NCCR SwissMAP (funded by the Swiss National Science Foundation, grant no. 205607), the Spanish Ministry of Science and Innovation MCIN/AEI/10.13039/ 501100011033 (CEX2019-000904-S and PID2020-113523GB-I00), the Spanish Ministry of Economic Affairs and Digital Transformation (project QUANTUM ENIA, as part of the Recovery, Transformation and Resilience Plan, funded by EU program NextGenerationEU), Comunidad de Madrid (QUITEMAD-CM P2018/TCS-4342), Universidad Complutense de Madrid (FEI-EU-22-06), and the CSIC Quantum Technologies Platform PTI-001.
\end{acknowledgments}

\bibliographystyle{quantum}
\bibliography{bibliography}

\appendix

\section{Nonlinear inequalities for chain networks}
\label{app:nlchain}

In this appendix we analyze the cases that are missing in Section \ref{sec:chainnl:long}, namely, $\ell$-QNN in chain networks with an even number of parties, and $\ell$-NN in chain networks with an arbitrary number of parties.
The core of the derivations is based on the concept of $k$-independent sets of sources \cite{Luo2018}.

\subsection{$\ell$-QNL bounds for even number of parties}
Let us begin discussing the case when the nonclassical sources distribute quantum systems.
An inequality for the case of chains with an odd number of parties was derived in Section \ref{sec:chainnl:long}, namely
\begin{equation*}
    I^{\frac{2}{n+1}}+J^{\frac{2}{n+1}}\leq 
    \begin{cases} 2^{\frac{n-1}{2(n+1)}} & \ell > \frac{n}{2} \\ \sqrt{2} & \ell<\frac{n}{2}
    \end{cases}.
\end{equation*}

For the case of even $n$, begin considering that the chain network contains $\ell$ classical sources with $\ell\geq n/2$.
Consider the two subnetworks $\cN_1$ and $\cN_2$, where $\cN_1$ contains the parties $\sA_{1}, \dots, \sA_{n-1}$, and $\cN_2$ contains the parties $\sA_{2}, \dots, \sA_{n}$.
Either subnetwork is a network with an odd number of parties which has, at least, $\ell-1$ classical sources.
From Eqs.~\eqref{DD1}-\eqref{DD5}, the correlations in $\cN_1$ and $\cN_2$ satisfy the inequality (\ref{DD5}) for $n-1$ parties, i.e., $I^{\frac{2}{n}}+J^{\frac{2}{n}}\leq 2^{\frac{n-2}{2n}}$.
One can thus test this inequality in the first $n-1$ parties and in the last $n-1$ parties to verify $\ell$-QNN with $\ell\geq n/2$.

If less than half of the sources in the network are classical, the situation is similar to that in Eq.~\eqref{DD6}.
Namely, the joint correlations generated by either $\cN_1$ or $\cN_2$ achieve the maximal bound of $\sqrt{2}$ for the inequality \eqref{DD5} with $n-1$ parties, i.e., $I^{\frac{2}{n}}+J^{\frac{2}{n}}\leq \sqrt{2}$.
This means that, in the case of an even number of parties and less than half of the sources being classical (i.e., for $\ell<n/2$), it is not currently possible to witness $\ell$-QNN.

\subsection{$\ell$-NN bounds}\label{app:lnnchain}
Now we focus on the bounds that are achieved when the only constraint on the nonclassical sources is that they distribute systems that satisfy the no-signaling principle.
As in the previous section, we distinguish the cases of networks with odd and even number of parties.

\textbf{Case 1}. Odd $n$ with $n\geq 3$.

Suppose that the chain network consists of $\ell$ classical variables with $\ell\geq 1$ while the other sources distribute arbitrary no-signaling systems, as shown in Fig.~\ref{fig:chains}(b).
Equation \eqref{DD1} holds for this kind of networks.
However, in contrast with the derivation in Section \ref{sec:chainnl:long}, for sources distributing nonsignaling systems the objects in Eq.~\eqref{DD3} are bounded by
\begin{equation}
    \begin{aligned}
        & \left|\left\langle \prod\limits_{{\rm{}odd}\, i;i\not\in S }A^+_{x_{i}}\prod\limits_{{\rm even}\, j}A_{x_{j}=0}\right\rangle\right|\leq 1,
        \\
        & \left|\left\langle \prod\limits_{{\rm{}odd}\, i;i\not\in S }A^{-}_{x_{i}}\prod\limits_{{\rm{}even}\, j}A_{x_{j}=1}\right\rangle\right|\leq 1.
    \end{aligned}
    \label{DD10}
\end{equation}
Take $n=5$ as an example, and $S=\{1\}$, i.e., that only the source connecting $\sA_1$ and $\sA_2$ distributes classical randomness.
Given a PR box, defined by $\textrm{PR}(a, b|x,y)= 1/2$ if $a \oplus b = xy$ and $0$ otherwise, and defining the total correlation as
\begin{equation*}
    \begin{aligned}
        P(a_2, \dots, &a_5|x_2,\dots, x_5)\\
        & = \textrm{PR}(a_{2},a_{3}|x_2,x_3)\textrm{PR}(a_{4},a_{5}|x_4,x_5),
        \end{aligned}
\end{equation*}
straightforward evaluation leads to saturating the upper bounds, i.e., $\left\langle A_{x_2=0}A^+_{x_3}A_{x_4=0}A^+_{x_5}\right\rangle=1$ and $\left\langle A_{x_2=1}A^-_{x_3}A_{x_4=1}A^-_{x_5}\right\rangle=1$.
Similar results hold for general $n\geq 3$.

Now, let $S$ consist of all odd integers $i$ such that $\sA_i$ shares classical variables with both $\sA_{i-1}$ and $\sA_{i+1}$.
Combining Eq.~\eqref{DD1} with the inequalities in Eq.~\eqref{DD10}, one finds
\begin{eqnarray}
    I^{\frac{2}{n+1}}+J^{\frac{2}{n+1}}
    &\leq &\left\langle \prod\limits_{i\in S}A^{+}_{x_{i}}\right\rangle^{\frac{2}{n+1}}+\left\langle \prod\limits_{i\in S}A^{-}_{x_{i}}\right\rangle^{\frac{2}{n+1}}
    \nonumber
    \\
    &\leq
    &\prod\limits_{i\in S}\cos^{\frac{4}{n+1}}\vta_i+\prod\limits_{i\in S}\sin^{\frac{4}{n+1}}\vta_i
    \label{DD12}
    \\
    &\leq &\sin^{\frac{4 |S|}{n+1}}\left(\frac{\pi}{2}-\frac{1}{|S|}\sum_{i\in S}\vta_i\right)
      \nonumber
    \\
    &&+ \sin^{\frac{4|S|}{n+1}}\left(\frac{1}{|S|}\sum_{i\in S}\vta_i\right)
    \label{DD13}
    \\
    &:=&\cos^{\frac{4|S|}{n+1}}\vta+\sin^{\frac{4|S|}{n+1}}\vta
    \label{DD14}
    \\
    &\leq & 2^{1-\frac{2|S|}{n+1}},
    \label{DD15}
\end{eqnarray}
where in Eq. \eqref{DD12} we have used that $\left|\left\langle A^{+}_{x_{i}}\right\rangle\right|\leq\cos^2\vta_i$ and $\left|\left\langle A^{-}_{x_{i}}\right\rangle\right|\leq\sin^2\vta_i$ for $\vta_i\in [0,\pi/2]$ as in the derivation of Eq.~\eqref{DD4}.
The inequality \eqref{DD13} is from \cite[Lemma 1]{Luo2018}, namely $\prod_{i=1}^k\sin^{\frac{1}{k}}\vta_i\leq \sin(k^{-1}\sum_{i=1}^k\vta_i)$, where $k=|S|$.
Note that, in this case, the relevant quantity for the bound is not the number of classical sources in the network, $\ell$, but the number of parties which only receive classical systems, $|S|$.
In Eq.~\eqref{DD14}, we have defined $\vta=|S|^{-1}\sum_{i\in S}\vta_i$.
The last inequality is from the fact that the unique extreme point of the continuous function $\cos^{\frac{4k}{n+1}}\vta+\sin^{\frac{4k}{n+1}}\vta$ for $\vta\in [0,\pi/2]$ is $\vta=\pi/4$.

\textbf{Case 2}. Even $n$ with $n\geq 4$.

As in the case in the previous subsection, a chain network with an even number of parties can be decomposed into subnetworks consisting of $n-1$ parties.
If there are $\ell$ classical sources, each subnetwork contains no less than $\ell-1$ classical sources, and thus the $n-1$-partite joint correlations satisfy the inequality \eqref{DD15} for $n-1$ parties.

\subsection{Quantum violations}\label{app:nlchain:quantum}
Now we prove that it is possible to violate Eq.~\eqref{nlchain} with realizations consisting of quantum states for models with $\ell\geq\frac{n-1}{2}$.
In quantum scenarios, suppose that each pair of parties, $\sA_i$ and $\sA_{i+1}$, shares one generalized EPR state defined in Eq.~\eqref{A006}, as shown in Fig.~\ref{fig:chains}(c).

For odd $n\geq 3$, Ref.~\cite{Luo2018} gave quantum observables $A_{x_1}, \dots, A_{x_n}$ such that
\begin{equation}
    I^{\frac{2}{n+1}}+J^{\frac{2}{n+1}}
    =\left(1+\prod\limits_{i=1}^{n-1}\sin^{\frac{4}{n+1}}2\ta_i\right)^{1/2}.
    \label{DD17}
\end{equation}
These measurements exceed the bound in Eq.~\eqref{DD5} whenever the $\ta_i$ satisfy $\Pi_{i=1}^{n-1}\sin2\ta_i >(2^{\frac{n-1}{n+1}}-1)^{\frac{n+1}{4}}$, thus certifying $\ell$-QNN for long chain quantum networks with $\ell \geq (n-1)/2$.
Moreover, as explained above, inequality \eqref{nlchain} cannot guarantee $\ell$-QNN of any quantum network if $\ell< (n-1)/2$.

For even $n$, consider two connected $n-1$-partite subnetworks $\cN_1$ and $\cN_2$, where $\cN_1$ consists of $\sA_{1}, \dots, \sA_{n-1}$ and $\cN_2$ consists of $\sA_{2}, \dots, \sA_{n}$.
The strategy of Ref.~\cite{Luo2018} gives $\ell$-QNN correlations in each subnetwork, since it also gives the value in Eq.~\eqref{DD17} for $\ell \geq n/2-1$.

Note that the bound in the inequality \eqref{DD15} is larger than the maximal quantum bound \cite{Luo2018} if $|S|\leq (n+1)/2$ for odd $n$.
From the definition of $S$ before Eq.~\eqref{DD12}, if there are at most $\lfloor (n-1)/4 \rfloor$ sources of no-signaling systems the bound in the inequality \eqref{DD15} is smaller than the bound in the inequality \eqref{nlchain}, where $\lfloor x \rfloor$ denotes the largest integer smaller than $x$.
This means that $\ell$-NN in chain networks with $\ell\geq n-\lfloor (n-1)/4 \rfloor=\lceil(3n-1)/4\rceil$ may be verified by violating the inequality \eqref{nlchain}, where $\lceil x \rceil$ denotes the smallest integer larger than $x$.
However, with our results it is not possible to verify FNN in any chain network if it has less than $\lceil(3n-1)/4\rceil$ classical sources.
This result can be extended for $n$-partite chain networks with even $n$, again, by using two $n-1$-partite connected subnetworks.

\section{Quantum violations of linear FNN inequalities for chain networks}
\label{app:lchain}
In this appendix we show quantum strategies that violate the inequalities \eqref{lbilocal} and \eqref{chainl}.

\subsection{The tripartite chain}
In quantum scenarios, consider the case where each source distributes generalized EPR states given by Eq.~\eqref{A006}.
The total state is thus given by
\begin{eqnarray}
    \ket{\Phi}&=&\cos\ta_1\cos\ta_2\ket{0000}
    +\cos\ta_1\sin\ta_2\ket{0011}
\nonumber
\\
&&
+\sin\ta_1\cos\ta_2\ket{1100}
    +\sin\ta_1\sin\ta_2\ket{1111}.
\end{eqnarray}

Let $A_0=\cos\vta \sigma_3+\sin\vta \sigma_1$, $A_1=\cos\vta \sigma_3-\sin\vta \sigma_1$, $B^{0}=\ketbra{\phi^+}{\phi^+}$, $B^{1}=\ketbra{\phi^-}{\phi^-}$, $B^{2}=\ketbra{\psi^+}{\psi^+}$, $B^{3}=\ketbra{\psi^-}{\psi^-}$, $C_0=\sigma_3$, $C_1=\sigma_1$, and $C_2=\sigma_1\sigma_3\sigma_1$, where $\ket{\phi^\pm}=\frac{1}{\sqrt{2}}(\ket{00}\pm \ket{11})$ and $\ket{\psi^\pm}=\frac{1}{\sqrt{2}}(\ket{01}\pm \ket{10})$.
After Bob's local measurement, the joint system shared by Alice and Charlie is, depending on Bob's outcome, one of the following states
\begin{equation*}
    \begin{aligned}
    &\ket{\Psi_0}=r_1(\cos\ta_1\cos\ta_2\ket{00}+\sin\ta_1\sin\ta_2\ket{11}),\\
    &\ket{\Psi_1}=r_1(\cos\ta_1\cos\ta_2\ket{00}-\sin\ta_1\sin\ta_2\ket{11}),\\
    &\ket{\Psi_2}=r_2(\cos\ta_1\sin\ta_2\ket{01}+\sin\ta_1\cos\ta_2\ket{10}),\\
    &\ket{\Psi_3}=r_2(\cos\ta_1\sin\ta_2\ket{01}-\sin\ta_1\cos\ta_2\ket{10}),
    \end{aligned}
\end{equation*}
with $r_1=1/\sqrt{\cos^2\ta_1\cos^2\ta_2+\sin^2\ta_1\sin^2\ta_2}$ and
$r_2=1/\sqrt{\cos^2\ta_1\sin^2\ta_2+\sin^2\ta_1\cos^2\ta_2}$.
Moreover, we have $p(b\,{=}\,0)=p(b\,{=}\,1)=1/(2r_1^2)$ and $p(b\,{=}\,2)=p(b\,{=}\,3)=1/(2r_2^2)$.

Note that $\cB_3$ can be written as a different CHSH game for each outcome of Bob.
Thus, we have that
\begin{eqnarray}
    \text{CHSH}_0&:=&\bra{\Psi_0}A_0C_0+A_0C_1+A_1C_0-A_1C_1\ket{\Psi_0}
    \nonumber\\
    &=&2\cos\vta+r_1^2\sin\vta\sin2\ta_1\sin2\ta_2,
    \label{Ee5}
    \\
    \text{CHSH}_1&:=&\bra{\Psi_1}A_0C_0-A_0C_1+A_1C_0+A_1C_1\ket{\Psi_1}
    \nonumber\\
    &=&2\cos\vta+r_1^2\sin\vta\sin2\ta_1\sin2\ta_2,\\
    \text{CHSH}_2&:=&\bra{\Psi_2}A_0C_1+A_0C_2-A_1C_1+A_1C_2\ket{\Psi_2}
    \nonumber\\
    &=&2\cos\vta+r_2^2\sin\vta\sin2\ta_1\sin2\ta_2,\\
    \text{CHSH}_3&:=&\bra{\Psi_3}-A_0C_1+A_0C_2+A_1C_1+A_1C_2\ket{\Psi_3}
    \nonumber\\
    &=&2\cos\vta+r_2^2\sin\vta\sin2\ta_1\sin2\ta_2.
    \label{Ee8}
\end{eqnarray}
Substituting Eqs.~\eqref{Ee5}-\eqref{Ee8} into Eq.~\eqref{lbilocal} gives
\begin{eqnarray*}
    \bra{\Phi}\cB_3\ket{\Phi}&=&\sum_ip(b=i)\text{CHSH}_i
    \nonumber
    \\
    &=&2(\cos\vta+\sin\vta \sin2\ta_1\sin2\ta_2).
\end{eqnarray*}
If, for instance, we choose $\vta$ such that it satisfies $\cos\vta=1/\sqrt{1+\sin^22\ta_1\sin^22\ta_2}$, we have that $\bra{\Phi}\cB_3\ket{\Phi}=2\sqrt{1+\sin^22\ta_1\sin^22\ta_2}$, therefore violating Eq.~\eqref{lbilocal} for any $\ta_1,\ta_2\in (0,\frac{\pi}{2})$.

\subsection{Long chains}
Consider the quantum realization in which all $n-1$ sources in the chain distribute states of the form \eqref{A006}.
The total state of the system is
\begin{equation*}
    \begin{aligned}
        \ket{\Phi}=&\sum_{i_1,\dot, i_n}\left(\prod_{j=1}^n\alpha_{i_j}\right)\ket{i_1,i_1i_2,\dots, i_{n-1}i_n,i_n}
        \\
        =&\sum_{i_2, \dots, i_n}\left[\left(\prod_{j=2}^{n-1}\alpha_{i_j}\right)\ket{0,0i_2,\dots, i_{n-1}0,0}\right.
        \nonumber
        \\
        &\qquad\quad+\left(\prod_{j=2}^{n-1}\alpha_{\oi_j}\right)\ket{1,1i_2,\dots, i_{n-1}1,1}
        \\
        &\qquad\quad+\left(\prod_{j=2}^{n-1}\alpha_{i_j}\right)\ket{0,0i_2,\dots, i_{n-1}1,1}
        \nonumber
        \\
        &\qquad\quad\left.+\left(\prod_{j=2}^{n-1}\alpha_{\oi_j}\right)\ket{1,1i_2,\dots, i_{n-1}0,0}\right],
    \end{aligned}
\end{equation*}
where $\oj=1\oplus j$, $\alpha_{i_j=0}=\cos\ta_j$ and $\alpha_{i_j=1}=\sin\ta_j$.
Moreover, define the quantum measurement observables
\begin{gather*}
   A^{(1)}_0=\cos\vta \sigma_3+\sin\vta \sigma_1,\, 
   A_1=\cos\vta \sigma_3-\sin\vta \sigma_1,
   \\
    M^{0,0}=\ketbra{\phi^+}{\phi^+},\quad
    M^{1,1}=\ketbra{\phi^-}{\phi^-},\\
    M^{0,1}=\ketbra{\psi^+}{\psi^+},\quad
    M^{1,0}=\ketbra{\psi^-}{\psi^-},\\
    A^{(n)}_0=\sigma_3,\, A^{(n)}_1=\sigma_1,\, A^{(n)}_2=\sigma_1\sigma_3\sigma_1.
    \end{gather*}
After all the projection measurements of $\sA_2,\dots, \sA_{n-1}$, each of which has a two-bit outcome $a_2,\dots, a_{n-1}\in \{00,\dots, 11\}$, the joint system of the particles held by $\sA_1$ and $\sA_n$ collapses into a state in one of the following four sets
\begin{widetext}
    \begin{equation*}
        \begin{aligned}
            &S_{00}:=\left\{r_{\vi}\left(
            \prod\limits_{j=1}^n\alpha_{i_j}\ket{00} + \prod\limits_{j=1}^n\alpha_{\overline{i}_j}\ket{11}\right) \forall a_2, \dots, a_{n-1}\, s.t.\, \oplus_{j=2}^{n-1}a_j=00\right\},
            \\
            &S_{01}:=\left\{r_{\vi}\left(
            \prod\limits_{j=1}^n\alpha_{i_j}\ket{01} + \prod\limits_{j=1}^n\alpha_{\overline{i}_j}\ket{10}\right) \forall a_2, \dots, a_{n-1}\, s.t.\, \oplus_{j=2}^{n-1}a_j=01\right\},
            \\
            &S_{10}:=\left\{r_{\vi}\left(
            \prod\limits_{j=1}^n\alpha_{i_j}\ket{01} - \prod\limits_{j=1}^n\alpha_{\overline{i}_j}\ket{10}\right) \forall a_2, \dots, a_{n-1}\, s.t.\, \oplus_{j=2}^{n-1}a_j=10\right\},
            \\
            &S_{11}:=\left\{r_{\vi}\left(
            \prod\limits_{j=1}^n\alpha_{i_j}\ket{00} -\prod\limits_{j=1}^n\alpha_{\overline{i}_j}\ket{11}\right)\forall a_2, \dots, a_{n-1}\, s.t.\, \oplus_{j=2}^{n-1}a_j=11\right\},
        \end{aligned}
    \end{equation*}
\end{widetext}
where $\oplus$ denotes the bit-wise XOR, and $r_{\vi}=1/\sqrt{\Pi_{j=1}^n\alpha_{i_j}^2+\Pi_{j=1}^n\alpha_{\overline{i}_j}^2}$.

Let $p_{ij}=\sum_{\oplus_{s=2}^{n-1}a_s=ij}p(a_2,\dots,a_{n-1})$, i.e., the total probability of all the measurements of $\sA_2,\dots, \sA_{n-1}$ such that the collapsed state is in the set $S_{ij}$. 
Similarly to Eqs.~\eqref{Ee5}-\eqref{Ee8}, for any state $\ket{\phi}$ in one subset $S_{0i}$ with $i=0,1$, there exists $\vta$ independent of $\ket{\phi}$ such that
\begin{equation}
    \begin{aligned}
        \text{CHSH}_{\ket{\phi}}\!:=\!&\left\langle\! A^{(1)}_0\!A^{(n)}_{2i}\!+\!A^{(1)}_0\!A^{(n)}_{2i+1}
         \!+\!A^{(1)}_1\!A^{(n)}_{2i}\!-\!A^{(1)}_1\!A^{(n)}_{2i+1}\!\right\rangle_{\!\phi}
         \\
        =&\,2\cos\vta+2\sin\vta r_{\vi}\prod_{j=1}^n\alpha_{i_j}
        \\
        \geq& \, 2\cos\vta+2\sin\vta\min_{\oplus_{i=1}^{n-1} a_i=0i}
        \left\{r_{\vi}\prod_{j=1}^n\alpha_{i_j}
        \right\}.
    \end{aligned}
    \label{E12a}
\end{equation}
Now, if we define $\vta$ such that $\sin\vta=\min_{\oplus_{i=1}^{n-1} a_i=0i}
\{1/\sqrt{1+r_{\vi}^2\Pi_{j=1}^n\alpha^2_{i_j}}\}$, it follows from the inequality \eqref{E12a} that 
\begin{equation*}
    \begin{aligned}
        \text{CHSH}_{\ket{\phi}}\geq  &\, 2\sqrt{1+\min_{\oplus_{i=1}^{n-1} a_i=0i}   
        \left\{ r_{\vi}^2\prod_{j=1}^n\alpha^2_{i_j}\right\}}
        \\
        =&\min_{\oplus_{i=1}^{n-1} a_i=0i}\left
         \{2\sqrt{1+r_{\vec{i}}^2\prod_{j=1}^n\alpha^2_{i_j}}\right\}.
    \end{aligned}
\end{equation*}

Similarly, for any state $\ket{\phi}$ in one subset $S_{1i}$, it follows the same result as above, just with $-\vta$ instead of $\vta$.
Substituting both in Eq.~\eqref{chainl} one has
\begin{align*}
\cB_n=\,&\sum\limits_{i,j=0,1}p_{ij}\text{CHSH}_{\ket{\phi}\in S_{ij}}
\\
\geq\, & 2(p_{00}\!+\!p_{10})\min\limits_{\oplus_{i=1}^{n-1} a_i=00}
\left\{
\sqrt{1+r_{\vi}^2\prod\limits_{j=1}^n\alpha^2_{i_j}}
\right\}
\nonumber
\\
&+2(p_{01}+p_{11})\min\limits_{\oplus_{i=1}^{n-1} a_i=01}
\left\{
\sqrt{1+r_{\vi}^2\prod\limits_{j=1}^n\alpha^2_{i_j}}
\right\}
\\
> \,&2\sum_{i,j=0,1}p_{ij}\\
=\,&2,
\end{align*}
violating Eq.~\eqref{chainl}.
We note that this result can be extended for general bipartite entangled pure states \cite{Gisin1991}, using the subspace spanned by two basis states following Ref.~\cite{Luo2018}.

\end{document}